\documentclass[pra,
amssymb,amsmath,amsmath,showpacs,reprint,twocolumn]{revtex4}
\usepackage{color}
\usepackage{graphicx}
\usepackage{epstopdf}
\usepackage{natbib}
\usepackage{wasysym}
\usepackage{hyperref}
\usepackage{dcolumn}

\hypersetup{%
   pdfpagemode=None, 
   pdfstartpage=1,
   pdfstartview=FitH,
   pdfmenubar=true,
   pdftoolbar=true,
   colorlinks = true,
   linkcolor=blue,
   citecolor=blue,
   bookmarksopen=false
 }

\newcommand{\Fkt}[1]{\,\mathsf {#1}}

\def\openone{\leavevmode\hbox{\small1\kern-3.3pt\normalsize1}}

\ifx\Tr\renewcommand{\Tr}{\Fkt{Tr}} 
\else\newcommand{\Tr}{\Fkt{Tr}}
\fi

\usepackage{booktabs}
\usepackage{graphicx}
\usepackage{amsmath}
\usepackage{indentfirst}
\begin{document}
\title{Analytical two-center integrals over Slater geminal functions}

\author{\sc Micha\l\ Lesiuk\footnote[1]{e-mail: lesiuk@tiger.chem.uw.edu.pl} 
and Robert Moszynski}
\affiliation{\sl Faculty of Chemistry, University of Warsaw\\
Pasteura 1, 02-093 Warsaw, Poland}
\date{\today}

\begin{abstract}
We present analytical formulas for the calculation of the two-center two-electron integrals in the basis of Slater
geminals
and products of Slater orbitals. Our derivation starts with establishing a inhomogeneous fourth-order ordinary
differential equation that is obeyed by the master integral, the simplest integral with inverse powers of all
interparticle distances. To solve this equation it was necessary to introduce a new family of special functions which
are defined through their series expansions around regular singular points of the differential equation. To increase the
power of the interparticle distances under the sign of the integral we developed a family of open-ended recursion
relations. A handful of special cases of the integrals is also analysed with some remarks on simplifications that
occur. Additionally, we present some numerical examples of the master integral that validate the usefulness and
correctness of the key equations derived in this paper. In particular, we compare our results with the calculations
based on the series
expansion of the $\exp(-\gamma r_{12})$ term in the master integral.
\end{abstract}

\maketitle

\section{Introduction}
\label{sec:intro}

It is a well-known fact since the landmark paper of Kato~\cite{kato57} that the exact eigenfunction $\Psi$ of
the Sch\"{o}dinger Hamiltonian must satisfy certain conditions at the coalescence points of the particles. These are
the so-called
\emph{cusp conditions}, expressed mathematically as:
\begin{align}
\lim_{r_{ij}\rightarrow0}\left(\frac{\partial \Psi}{\partial
r_{ij}}\right)_{av}=\mu_{ij}q_iq_j\Psi(r_{ij}=0)
\end{align}
where $q_i$ are the charges of the particles, $\mu_{ij}$ is the reduced mass of the particles $i$ and $j$, and the
subscript $av$
denotes the spherical average over an infinitesimal sphere around $r_{ij}=0$. The above constraint must be satisfied
for every single pair of particles in the system. While the nuclear cusp condition is naturally satisfied by the
one-electron basis constructed from the Slater orbitals, electronic cusp condition appears to be a far more difficult
problem. Hill~\cite{hill85} analysed a simple example of a two-electron one-center system with the basis set taken as
the partial wave expansion:
\begin{align}
\begin{split}
\Psi(\vec{r}_1,\vec{r}_2)&=\sum_{lm}^LY_{lm}(\theta_1,\varphi_1)Y_{l,-m}(\theta_2,\varphi_2)\\
&\times \sum_{nn'}^Nf_{nlm}(r_1)f_{n'lm }(r_2)
\end{split}
\end{align}
where $Y_{lm}$ are spherical harmonics, $(\theta_i,\varphi_i)$, $i=1,2$, are the spherical angles of the vector
$\vec{r}_i$, and $f_n$ are some radial factors. He found that the error of the energy decays
as $\sim(L+1)^{-3}$, so a rather slow convergence is obtained. This sad corollary can be attributed to the fact that
the partial wave expansion has severe difficulties in fulfilling the electronic cusp condition. Much faster convergence
can be expected when the basis set is extended to include the $r_{12}$ factor explicitly. The latter finding is a
theoretical underpinning for a vast family of the so-called \emph{explicitly correlated} methods.

Explicitly correlated calculations in quantum mechanics have a long history. The first calculations of
this type were performed on the $1^1S$ state of the helium atom by Hylleraas in his classical 1929
paper~\cite{hylleraas27}. The Hylleraas
Ansatz for the wave function of He ground state was:
\begin{align}
\Psi_N=e^{-\zeta s}\sum_k^N c_k s^{l_k} t^{2 m_k}u^{n_k}
\end{align}
where $s=r_1+r_2$, $t=r_1-r_2$, $u=r_{12}$, and $r_i$ are the coordinates of electrons. Using a six term wave function
of
the above form with one nonlinear parameter Hylleraas obtained a result with three correct significant
digits in the ionization energy of helium~\cite{hylleraas27}. The length of this expansion can be increased
and it is a relatively
easy task to obtain a nanohartree accuracy. Many authors tried to extend the form of the Hylleraas Ansatz. For
instance,
Kinoshita~\cite{kinoshita57,kinoshita59} suggested to include negative powers of $s$ and $u$, and
Schwartz~\cite{schwartz60,schwartz62,schwartz63} included half-integer powers of the latter quantities. Several
researchers~\cite{bartlett37,frank66} included logarithmic terms \emph{e.g.} $\mbox{Log}(s)$ in order to satisfy
the three-particle coalescence condition of both the electrons and the nucleus. Further extension can be done by
considering
so-called ``double basis set'' ~\cite{drake88,drake94,drake99} in which each combination of powers of $r_1$, $r_2$, and
$r_{12}$ is included twice, but with different exponential scale factors and no logarithmic terms. Probably the most
well-known calculations in this basis set are those of Drake \emph{et al.}~\cite{drake02}, where about twenty
significant digits accuracy on the energy was reached. Of course, this idea can further be extended to the ``triple
basis
set'' and so forth. Important from the point of view of the present paper is the work of Korobov~\cite{korobov02}
who obtained a $25$ significant digits accuracy by using Slater-type geminals, \emph{i.e.} the wave function expanded as
a linear combination of the functions:
\begin{align}
\phi_k=e^{-\alpha_k r_1-\beta_k r_2-\gamma_k r_{12}}
\end{align}
where $\alpha_k$, $\beta_k$, $\gamma_k$ are complex parameters which were generated quasirandomly. Recently, Nakashima
and Nakatsuji used a method called iterative complement interaction (ICI), described in Ref.~\cite{nakashima07}, and
obtained forty significant digits accuracy which is the highest available until now. At the end of this short survey
over the helium atom we must admit that the exponentially correlated Gaussian functions (ECG) were also used with
success, see Ref.~\cite{pachucki04}. Of course, all of the methodologies mentioned above can equally well be applied to
the excited states and properties~\cite{lach04,moszynski87} of the He atom, and
and its isoelectronic series such as H$^-$ or Li$^+$. These systems were also
subjects of intensive studies in the past~\cite{luchow94,king97,yan98,thakkar02}.

The first explicitly correlated calculations on a molecular system, the hydrogen molecule, were carried out in 1933 by
James and Coolidge~\cite{james33} with a basis set, named today after them (JC), of the form:
\begin{align}
\xi_1^k\eta_1^l\xi_2^m\eta_2^n r_{12}^\mu e^{-\alpha \xi_1-\beta \xi_2}
\end{align}
where $\xi_i$ and $\eta_i$ are elliptical coordinates. In the advent of computers Ko\l os and Roothaan used this basis
set
to obtain a microhartree accuracy in the energy calculations~\cite{kolos60a,kolos60b}. Later on, Ko\l os and Wolniewicz
extended
the form of the above basis set to include the Heitler-London function, thereby allowing to describe
the dissociation of the molecule properly~\cite{kolos65}. It gave rise to so-called Ko\l os-Wolniewicz
(KW) basis set. 
The approaches described above were subsequently extended to the excited states of H$_2$ \emph{cf.}
Refs.~\cite{kolos90,frye89,liu94}. During the past decades several authors reported
calculations in the JC~\cite{kolos64,bishop78,sims06} or KW~\cite{kolos78,kolos86,wolniewicz93,wolniewicz95,kolos94}
basis
sets with an increasing accuracy. Among other approaches ICI calculations
presented by Nakatsuji \emph{et al.}~\cite{nakatsuji07} are worth noticing. It is rather astonishing that in the field
of H$_2$ ECG calculations were
proven to be very successful and even competitive with the approaches based on Slater
functions~\cite{jeziorski79,rychlewski94,cencek96,cencek08}. Recently, Pachucki, in his \emph{tour de force} paper,
derived
analytical equations for the integrals over the JC basis set~\cite{pachucki09}. This allowed to perform calculations
on
H$_2$ 
with at least fifteen digits accuracy, the highest accuracy
reported until now~\cite{pachucki10}. Let us end this paragraph by remarking that the two-electron analogues of H$_2$,
HeH$^+$~\cite{kolos76a,kolos76b,bishop79,cencek95,pachucki12} and He$_2^{2+}$~\cite{pauling33,yasigawa77,wolniewicz99},
were also studied in the literature.

The lithium atom and three-electron ions are probably the last example when Hylleraas-type basis set could still
successfully be
applied. It was possible because analytical equations for the resulting
integrals~\cite{fromm87,remiddi91,harris97} and useful recursion relations~\cite{pachucki04a,harris09} between them are
all known. This allowed very accurate calculations, among which those of King~\cite{king95}, Yan \emph{et
al.}~\cite{yan95,yan98}, and Puchalski and Pachucki~\cite{puchalski06} should be mentioned. The results of Hylleraas-CI
and
ECG calculations for the lithium atom are also available~\cite{luchow94}.
The accuracy of the calculations for the lithium atom cannot compete with that for helium. Nevertheless, the reported
energy
values still agree excellently with the best available experimental data~\cite{puchalski06}. The applicability of the
explicitly correlated calculations with the Hylleraas-like Ansatz is narrowed dramatically when passing to many-center
and/or many-electron systems. Since the Hylleraas-CI and ECG are the only methods that can be used in practice
for systems such as beryllium atom~\cite{schwegler94,komasa84,busse98}, the accuracy deteriorates significantly. Similar
situation holds for other few-body systems, H$_3^+$~\cite{cencek95,cencek98,frye90},
H$_3$~\cite{cencek93,komasa96}, He$_2$~\cite{komasa01}, and LiH \cite{cencek00}.

For many-electron systems explicitly-correlated variational
calculations are not feasible at the present. This is due to the high complexity in the
space and permutational symmetry of the wave function. However, basis functions including
the explicit dependence on the interelectronic distance $r_{12}$ can be introduced into
the many-body theory of many-electron systems. Indeed, it was realised as early as in
1966 by Byron and Joachain~\cite{byron66,byron67} and later by Pan and King~\cite{pan70,pan72}, Jeziorski, Szalewicz,
and collaborators~\cite{chalas77,szalewicz79,szalewicz82,szalewicz83a,szalewicz83b,szalewicz84} and Adamowicz and
Sadlej~\cite{adamowicz77,adamowicz78a,adamowicz78b} that the pair functions
appearing in the energy expressions of the many-body perturbation theory (MBPT), also known as the M\o{}ller-Plesset 
perturbation theory, can be expanded in terms of explicitly correlated functions,
provided that the strong orthogonality condition is satisfied. Since the strong orthogonality
condition is difficult to meet, Szalewicz {\em et al.}~\cite{szalewicz82,szalewicz83a,szalewicz83b,szalewicz84}
suggested to weaken it without
loosing the mathematical correctness of the theory. These early explicitly correlated MBPT approaches
employed the Hylleraas basis in the case of calculations of Byron and Joachain~\cite{byron66,byron67}
on the beryllium atom, and explicitly correlated Gaussian functions in case of the calculations
on the Be, LiH, Ne, and H$_2$O systems~\cite{wenzel86,bukowski99,bukowski98,jeziorska88}. In the early 1980's 
explicitly-correlated Gaussian geminals were used with success by Jeziorski and Szalewicz in
the coupled cluster (CC) calculations~\cite{szalewicz84}.
One important drawback of the approach summarized above is that the perturbation theory
and coupled cluster calculations involving explicitly correlated basis functions require
calculations of three and in some cases four-electron integrals. This makes this kind of
calculations prohibitively expensive and limited to small systems. A breakthrough in this
respect was suggested by Klopper and Kutzelnigg~\cite{kutz85,klopper87,klopper90}
for the MBPT calculations and by Noga and collaborators~\cite{noga94,noga97} for the
CC calculations. These authors suggested to include only terms linear in
the interelectronic distance $r_{12}$ and use an approximate resolution of identity to
approximate many-electron integrals with the two-electron integrals. In this way the
problem of calculating many-electron integrals was eliminated, although only in an
approximate way. Still, this approach was shown to be very successful in many 
spectroscopic and chemical applications. See, for instance, Ref.~\cite{klopper12} for a review.
Finally, the most recent advance in this field are the so-called explicitly-correlated CC-F12
methods~\cite{skhv08a,skhv08b,krt08,tew07,bokhan08,tew08}, in which
the interelectronic distance, $r_{12}$, is explicitly introduced into the pair functions through the exponential
correlation factor $\exp(-\gamma r_{12})$. The F12 methods have recently
been implemented in an efficient manner~\cite{hattig10,adler07,werner10} and shown to accelerate
the convergence towards the basis-set limit for a number of properties~\cite{neiss07,yang09a,yang09b}.
Unfortunately, the F12 method fails to reproduce accurate interaction potentials of diatomic molecules
\cite{patk1,patk2}, although it was shown to work well in the Li+LiH case \cite{skomorowski11}. 

In this paper we introduce a new basis set for accurate calculations on diatomic molecules,
the basis of Slater geminals. This basis can be used both in the variational calculations 
and in the many-body MBPT/CC theories. The Slater geminal basis has several advantages over
the explicitly correlated basis sets used in molecular calculations thus far. Among others, 
it satifies both the electron-nuclei and electron-electron cusp conditions. Similarly as for 
atoms, the exponential correlation factor is expected to improve the convergence of the 
short-range correlations, while the Slater type one-electron part will greatly reduce
the size of the expansion, thus leading to results much more accurate than possible at
present. This is especially important for the new emerging field at the border of
chemistry and physics, ultracold molecules. See the 2012 special issue of Chemical 
Reviews, and in particular papers by Quemener and Julienne \cite{julienne12},
Weidem\"uller and collaborators \cite{weidemuller12}, and by Koch and Shapiro \cite{koch12}.
To better appreciate the importance of high quality basis sets for molecular
calculations on diatomic molecules, let us just quote calculations on the
Sr$_2$ molecule \cite{skomorowski12a,skomorowski12b}, which are currently used in the
interpretation of the experimental data for the determination of the
time variation of the electron to proton mass ratio \cite{zelevinsky08,zelevinsky12}.
Another very appealing application of the Slater geminals for diatomic molecules are the
calculations of the relativistic effects. Indeed, when the relativistic corrections are 
calculated in the framework of the perturbation method and with the Breit-Pauli Hamiltonian 
it is necessary to calculate integrals with the $1/r_{12}^2$ factor.
Analytical calculation of such integrals in the two-center case was impossible until now. It was
necessary to use the infinite expansion in the  Gegenbauer polynomials, according to the scheme 
advocated by Wolniewicz~\cite{wolniewicz93}. Using our analytical equations for the integrals over the 
Slater geminal basis, all the necessary relativistic integrals involving the $1/r_{12}^2$ factor are obtained by a
simple one-dimensional 
numerical integration. Similar scheme was recently successfully applied to calculation of the 
relativistic corrections for the lithium atom~\cite{puchalski06}.

The paper is organised as follows. In Sec. II we define the master integral, $f(r)$, which will
serve as a generating integral for the calculations of all the integrals from the family 
(\ref{eq0a}) and derive a differential equation satisfied by $f(r)$. In Sec. III we show how solve the
homogeneous differential equation, thereby involving a new family of special functions. In Sec. IV
we derive solutions of the inhomogeneous differential equation so that an analytical expression for the master
integral becomes known explicitly. In Sec. V we establish a family of recursion relations that allow
calculations of the integrals with arbitrary powers of all electron-nuclear distances. Similar
procedure is adopted in Sec. VI to let arbitrarily grow the power of $r_{12}$ in the integrals. In
Sec. VII we consider a handful of special cases of the integrals that cannot be calculated with
the results of the previous sections. In these special cases, an analytical equation for the master integral
is found in terms of well-known special functions. In Sec. VIII we present some numerical examples
of the master integral that validate the usefulness and correctness of the analytical equations
derived in this paper. In particular, we compare our results with the calculations based on the series
expansion of the $\exp(-\gamma r_{12})$ term in the master integral. Finally, in Sec. IX we conclude our paper.

In the paper we highly rely on the known special functions to simplify the derivation and the final formulas. Our
convention for all special functions appearing below is the same as in Ref.~\cite{stegun72}. We also use Meijer
$G$-function which is defined according to Ref.~\cite{adrews85}.

\section{The master integral}
\label{sec:master}

In this paper, we consider analytical calculation of the two-electron integrals in the basis of
Slater geminals and Slater functions for a diatomic molecule. The latter basis set has the
general form:
\begin{align}
\label{eq0}
\begin{split}
\phi(\vec{r}_1,\vec{r}_2)&=
r_{1A}^ir_{1B}^jr_{2A}^kr_{2B}^lr_{12}^n\\
&\times e^{-u_3 r_{1A}-u_2 r_{1B}-w_2 r_{2A}-w_3 r_{2B}-w_1 r_{12}},
\end{split}
\end{align}
so it gives rise to the following class of two-electron two-center integrals:
\begin{align}
\label{eq0a}
\begin{split}
&f_n(i,j,k,l;u_2,u_3,w_2,w_3,w_1)=\\
&\int d^3 r_1\int d^3 r_2\;
r_{1A}^i r_{1B}^j r_{2A}^k r_{2B}^l r_{12}^n\\
&\times e^{-u_3 r_{1A}-u_2 r_{1B}-w_2 r_{2A}-w_3 r_{2B}-w_1 r_{12}},
\end{split}
\end{align}
where we adopted the following notation: $\vec{r}_i$, $i=1,2$,
denotes the coordinates of the electrons and $\vec{r}_K$, $K=A,B$, denotes the coordinates of
the nuclei. Consequently, $r_{iK}=|\vec{r}_{i}-\vec{r}_K|$ and $r_{12}=|\vec{r}_1-\vec{r}_2|$
denote the electron-nucleus and interelectronic distances, respectively. The above notation will be used
throughout the paper. 

It is noteworthy that the requirement $u_2>0$, $u_3>0$, $w_2>0$, $w_3>0$, and $w_1>0$ is
sufficient but much too strong to make the functions (\ref{eq0}) square-integrable. This requirement can
be significantly weakened by demanding only $u_2+u_3+w_1>0$ and $w_2+w_3+w_1>0$. Therefore, some of the nonlinear
parameters can be negative without violation of the square-integrability principle. This result is reminiscent of the
three-body Hylleraas integrals which will be discussed later. 

If the basis set is chosen in terms of spherical
harmonics multiplied by the radial factor and the exponential correlation factor, then using simple
manipulations based on the ordinary trigonometric relations, one can express the resulting integrals
in terms of combinations of the integrals from the family (\ref{eq0a}).

When performing calculations for a two-electron and diatomic system described by the Schr\"{o}dinger
Hamiltonian in the basis set defined by Eq. (\ref{eq0}), all the matrix elements of the operators 
are readily expressed through the integrals (\ref{eq0a}) except for the kinetic energy operator.
To express the latter quantities through the combinations of the integrals from the family (\ref{eq0a}), a somehow long
derivation is required. Not to disturb the consistency of the paper, this derivation is reported in the Appendix
\ref{app:appa}. As a
result, the matrix elements of the Schr\"{o}dinger Hamiltonian and all the integrals appearing in the nonrelativistic
molecular physics in the basis (\ref{eq0}) are expressed
fully analytically.

\subsection{Definition and the momentum space representation}
\label{subsec:def}

The master integral is defined as the simplest two-electron integral with inverse powers of all
electron-nuclear and
interelectronic distances, namely:
\begin{align}
\label{eq1}
\begin{split}
f(r)&=\int \frac{d^3 r_1}{4\pi}\int \frac{d^3 r_2}{4\pi}
\frac{e^{-u_3 \,r_{1A}}}{r_{1A}} \frac{e^{-u_2 \,r_{1B}}}{r_{1B}} \\
&\times\frac{e^{-w_2 \,r_{2A}}}{r_{2A}} 
\frac{e^{-w_3 \,r_{2B}}}{r_{2B}} \frac{e^{-w_1 \,r_{12}}}{r_{12}} r,
\end{split}
\end{align}
where the notation for all appearing quantities is the same as in Eq. (\ref{eq0}) and $r=r_{AB}$ is the internuclear
distance. The reason for
the choice of the multiplicative constant $\frac{r}{(4\pi)^2}$ and the particular notation for the
nonlinear parameters will be clear from the further derivation. Once this integral is
known analytically, all integrals $f_n$ of Eq. (\ref{eq0a}) can be obtained by multiple
differentiations of Eq. (\ref{eq1}) over the nonlinear parameters $u_2$, $u_3$, $w_2$, $w_3$, and $w_1$.

Our first task is to derive an analytical equation for the above integral. We perform a Laplace transform of the
master integral with respect to $r$ and therefore define another integral $g(u_1)$:
\begin{align}
\label{eq2}
\begin{split}
g(u_1)&=\int_0^\infty dr f(r) e^{-u_1\,r}=\int \frac{d^3 r}{4\pi}\frac{f(r)}{r^2}e^{-u_1\,r}.
\end{split}
\end{align}
This equality allows us to calculate $f(r)$ from the inverse Laplace transform formula:
integral
\begin{align}
\label{eq3}
f(r)=\frac{1}{2\pi\dot{\imath}}\int_{-\dot{\imath}\infty+\epsilon}^{\dot{\imath}
\infty+\epsilon} du_1\; g(u_1) e^{u_1\,r}.
\end{align}
The explicit form of the integral $g(u_1)$ can conveniently be written, after the simple interchange of variables
$\vec{\rho}_1=\vec{r}_{12}$, $\vec{\rho}_2=\vec{r}_{2A}$, $\vec{\rho}_3=\vec{r}_{2B}$, as:
\begin{align}
\label{eq4}
\begin{split}
g(u_1)&=\int\frac{d^3 \rho_1}{4\pi}\int\frac{d^3 \rho_2}{4\pi}\int\frac{d^3 \rho_3}{4\pi}\;
\frac{e^{-u_3 \,\rho_{12}}}{\rho_{12}} \frac{e^{-u_2 \,\rho_{31}}}{\rho_{31}}\\
&\times\frac{e^{-w_2 \,\rho_2}}{\rho_2} 
\frac{e^{-w_3 \,\rho_3}}{\rho_3} \frac{e^{-w_1 \,\rho_1}}{\rho_1} \frac{e^{-u_1
\,\rho_{23}}}{\rho_{23}},
\end{split}
\end{align}
with $\rho_{12}=|\vec{\rho}_1-\vec{\rho}_2|$ and analogous formulas for $\rho_{13}$ and $\rho_{23}$.
The above representation is familiar as it is the generating integral from the theory of three-electron one-center
integrals~\cite{fromm87,remiddi91}. Let us recall the momentum space representation of $g(u_1)$:
\begin{align}
\label{eq4a}
g(u_1)=G(1,1,1,1,1,1),
\end{align}
where
\begin{align}
\label{eq5}
\begin{split}
&G(m_1,m_2,m_3,m_4,m_5,m_6)= \\
&\frac{1}{8\pi^6}\int d^3 k_1 \int d^3 k_2 \int d^3 k_3
\frac{1}{(k_1^2+u_1^2)^{m_1}}
\frac{1}{(k_2^2+u_2^2)^{m_2}} \\
&\times\frac{1}{(k_3^2+u_3^2)^{m_3}} 
\frac{1}{(k_{32}^2+w_1^2)^{m_4}}
\frac{1}{(k_{13}^2+w_2^2)^{m_5}}
\frac{1}{(k_{21}^2+w_3^2)^{m_6}}.
\end{split}
\end{align}

\subsection{Differential equation in the momentum space}
\label{subsec:mom}

In this subsection we establish a differential equation for $G(1,1,1,1,1,1)$. Let us first denote the integrand in Eq.
(\ref{eq5}) by $\tilde{G}$ with an analogous notation for its parameters:
\begin{align}
\label{eq8}
\begin{split}
\tilde{G}&(m_1,m_2,m_3,m_4,m_5,m_6)=\\
&\frac{1}{(k_1^2+u_1^2)^{m_1}}
\frac{1}{(k_2^2+u_2^2)^{m_2}} 
\frac{1}{(k_3^2+u_3^2)^{m_3}} \\
\times &\frac{1}{(k_{32}^2+w_1^2)^{m_4}}
\frac{1}{(k_{13}^2+w_2^2)^{m_5}}
\frac{1}{(k_{21}^2+w_3^2)^{m_6}}.
\end{split}
\end{align}
Our derivation is based on the so-called \emph{integration by parts identities}~\cite{tkachov81,chetyrkin81} and
the fact that due to Green theorem the following family of integrals vanish:
\begin{align}
\label{eq9}
\begin{split}
0&=I_{ij}\\
&=\frac{1}{8\pi^6}\int d^3 k_1 \int d^3 k_2 \int d^3 k_3
\vec{\nabla}_j \cdot \left[ \vec{k}_i \tilde{G}(1,1,1,1,1,1) \right],
\end{split}
\end{align}
where the $i$ and $j$ indices can independently take values $1,2$, and $3$. The above identity provides nine
equations that relate the values of $G$ with different arguments. These equations can be divided into
three sets, the  first set being $I_{13}, I_{23}, I_{33}$ and the two other obtained by a
permutation of the second index. It can be proven that to derive the desired differential equation
only one of these sets has to be considered and the results from the others are identical. Therefore,
we will consider the trio $I_{13}, I_{23}, I_{33}$ but this choice is arbitrary. To give
an example we will show the derivation for $I_{13}$. It follows from the definition that:
\begin{align}
\label{eq10}
\begin{split}
&\vec{\nabla}_3\cdot \left[ \vec{k}_1 \tilde{G}(1,1,1,1,1,1) \right]=\\
&-2\vec{k}_1\cdot \vec{k}_3 \tilde{G}(1,1,2,1,1,1)
-2\vec{k}_1\cdot \vec{k}_{32} \tilde{G}(1,1,1,2,1,1)\\
&+2\vec{k}_1\cdot \vec{k}_{13} \tilde{G}(1,1,1,1,2,1).
\end{split}
\end{align}
The scalar (dot) products of several $\vec{k}$ vectors appearing in the above equation are expanded using the relation
$\vec{k}_1\cdot \vec{k}_3=-\frac{1}{2}\left[ \vec{k}_{13}^2 - \vec{k}_{1}^2 - \vec{k}_{3}^2 \right]$ and similar for
other possible combinations. This allows to rewrite the r.h.s. of Eq. (\ref{eq10}) as:
\begin{align}
\label{eq12}
\begin{split}
&\left[ \vec{k}_{13}^2 - \vec{k}_{1}^2 - \vec{k}_{3}^2 \right]\tilde{G}(1,1,2,1,1,1)\\
&+\left[ \vec{k}_{13}^2 - \vec{k}_{12}^2 - \vec{k}_{3}^2 + \vec{k}_{1}^2
\right]\tilde{G}(1,1,1,2,1,1)\\
&+\left[\vec{k}_{13}^2 - \vec{k}_{3}^2 -\vec{k}_1^2 \right] \tilde{G}(1,1,1,1,2,1).
\end{split}
\end{align}
The next step is to make all the coefficients multiplying the different $\tilde{G}$ functions independent
of the $\vec{k}$ vectors. The latter are absorbed into $\tilde{G}$ in the following way:
\begin{align}
\label{eq13}
\vec{k}_{1}^2 \tilde{G}(1,1,2,1,1,1)=\tilde{G}(0,1,2,1,1,1)-u_1^2 \tilde{G}(1,1,2,1,1,1).
\end{align}
After necessary simplifications the expression for $I_{13}$ becomes:
\begin{align}
\label{eq14}
\begin{split}
I_{13}=&(u_3^2+u_1^2-w_2^2)G(1,1,2,1,1,1)\\
+&(w_3^2+u_3^2-u_2^2-w_2^2)G(1,1,1,2,1,1)\\
+&(u_3^2-u_1^2-w_2^2)G(1,1,1,1,2,1)+G(1,1,2,1,0,1)\\
-&G(0,1,2,1,1,1)+G(1,1,1,2,0,1)-G(1,1,1,2,1,0)\\
-&G(1,1,0,2,1,1)+G(1,0,1,2,1,1)-G(1,1,0,1,2,1)\\
+&G(0,1,1,1,2,1).
\end{split}
\end{align}
In a very similar way the expressions for $I_{23}$ and $I_{33}$ can be derived. The final equations are:
\begin{align}
\label{eq15}
\begin{split}
I_{23}=&(u_2^2+u_3^2-w_1^2)G(1,1,2,1,1,1)\\
+&(w_3^2+u_3^2-u_1^2-w_1^2)G(1,1,1,1,2,1)\\
+&(u_3^2-w_1^2-u_2^2)G(1,1,1,2,1,1)+G(1,0,1,2,1,1)\\
-&G(1,1,0,1,2,1)+G(0,1,1,1,2,1)-G(1,0,2,1,1,0)\\
-&G(1,1,0,2,1,1)+G(1,1,2,0,1,1)-G(1,1,1,1,2,0)\\
+&G(1,1,1,0,2,1),
\end{split}
\end{align}
\begin{align}
\label{eq16}
\begin{split}
I_{33}=&2u_3^2G(1,1,2,1,1,1)+(w_2^2-u_1^2+u_3^2)G(1,1,1,1,2,1)\\
+&(u_3^2+w_1^2-u_2^2)G(1,1,1,2,1,1)+G(1,0,1,2,1,1)\\
-&G(1,1,1,1,1,1)+G(0,1,1,1,2,1)\\
-&G(1,1,0,1,2,1)-G(1,1,0,2,1,1).
\end{split}
\end{align}
By an inspection of these three equations we note that all the $G$ integrals fall
into three classes. The first class consists of integrals with one of the parameters $m_1,...,m_6$
equal to zero. It is easy to verify by a direct calculation that these integrals belong to the class
of the well-known Hylleraas-type helium (three-body) integrals:
\begin{align}
\label{eq17}
\begin{split}
\Gamma\left(n_1,n_2,n_3;\alpha,\beta,\gamma\right)&=
\int \frac{d^3 r_1}{4\pi} \int \frac{d^3 r_2}{4\pi}
r_1^{n_1-1}r_2^{n_2-1}r_{12}^{n_3-1}\\
&\times e^{-\alpha r_1-\beta r_2-\gamma r_{12}},
\end{split}
\end{align}
and analytical equations for these integrals are all known since they can be obtained from the generating
integral:
\begin{align}
\Gamma\left(0,0,0;\alpha,\beta,\gamma\right)=
\frac{1}{(\alpha+\beta)(\alpha+\gamma)(\beta+\gamma)},
\end{align}
by a proper differentiation or integration with respect to the nonlinear parameters $\alpha, \beta,
\gamma$. Recursion relations that enable generation of $\Gamma$ with arbitrary values of $n_1,n_2,n_3$ were presented
long time ago by Ko\l os and co-workers~\cite{sack67}. An analytical expression to generate the integral
$\Gamma(0,0,0)$ was derived earlier~\cite{calais62}. 

The second class of integrals consists of $G(1,1,2,1,1,1),
G(1,1,1,2,1,1)$, and $G(1,1,1,1,2,1)$, and the third class is the master integral $G(1,1,1,1,1,1)$.
Therefore, we solve the set of three equations (\ref{eq14})$-$(\ref{eq16}) with respect to one of the integrals from the
second class. Let us choose $G(1,1,1,2,1,1)$. The result is:
\begin{align}
\label{eq18}
\begin{split}
&\frac{1}{2}\frac{\partial \sigma}{\partial w_1}G(1,1,1,1,1,1)-2w_1 \sigma G(1,1,1,2,1,1)\\
&+P(w_1,u_1;w_2,u_2;w_3,u_3)=0,
\end{split}
\end{align}
where $\sigma$ is a polynomial in all nonlinear parameters:
\begin{align}
\label{eq19}
\begin{split}
\sigma=&u_1^2w_1^2\left(u_1^2-u_2^2-u_3^2+w_1^2-w_2^2-w_3^2\right)\\
+&u_2^2w_2^2\left(-u_1^2+u_2^2-u_3^2-w_1^2+w_2^2-w_3^2\right)\\
+&u_3^2w_3^2\left(-u_1^2-u_2^2+u_3^2-w_1^2-w_2^2+w_3^2\right)\\
+&u_1^2u_2^2 w_3^2+u_1^2 u_3^2 w_2^2+u_2^2u_3^2w_1^2+w_1^2w_2^2 w_3^2.
\end{split}
\end{align}
The function $P(w_1,u_1;w_2,u_2;w_3,u_3)$ is a combination of integrals from the first class
with coefficients being some polynomials in the nonlinear parameters. Its derivation is long and does
not present any advance over already published formulas~\cite{pachucki09,puchalski10}, so we list here only the
final equation:
\begin{widetext}
\begin{align}
\label{eq19a}
\begin{split}
&P(w_1,u_1;w_2,u_2;w_3,u_3)= \\
&=-u_1w_1\left[(u_1+w_2)^2-u_3^2\right]\Gamma(0,0,-1;u_1+w_2,u_3,u_2+w_1)  \\
&-u_1w_1\left[(u_1+u_3)^2-w_2^2\right]\Gamma(0,0,-1;u_1+u_3,w_2,w_1+w_3)  \\
&+\left[u_1^2w_1^2+u_2^2w_2^2-u_3^2w_3^2+w_1w_2(u_1^2+u_2^2-w_3^2)
\right]\Gamma(0,0,-1;w_1+w_2,w_3,u_1+u_2)  \\
&+\left[u_1^2w_1^2-u_2^2w_2^2+u_3^2w_3^2+w_1w_3(u_1^2+u_3^2-w_2^2)
\right]\Gamma(0,0,-1;w_1+w_3,w_2,u_1+u_3)  \\
&-\left[u_2(u_2+w_1)(u_1^2+u_3^2-w_2^2)-u_3^2(u_1^2+u_2^2-w_3^2) \right]\Gamma(0,0,-1;u_2+w_1,u_3,u_1+w_2)
 \\
&-\left[u_3(u_3+w_1)(u_1^2+u_2^2-w_3^2)-u_2^2(u_1^2+u_3^2-w_2^2) \right]\Gamma(0,0,-1;u_3+w_1,u_2,u_1+w_3)
 \\
&+w_1\left[w_2(u_1^2-u_2^2+w_3^2)+w_3(u_1^2+w_2^2-u_3^2)\right]\Gamma(0,0,-1;w_2+w_3,w_1,u_2+u_3)  \\
&+w_1\left[u_2(u_1^2-w_2^2+u_3^2)+u_3(u_1^2+u_2^2-w_3^2)\right]\Gamma(0,0,-1;u_2+u_3,w_1,w_2+w_3),
\end{split}
\end{align}
\end{widetext}
where
\begin{align}
\Gamma(0,0,-1;\alpha,\beta,\gamma)=\frac{\mbox{Log}\left( \frac{\gamma+\alpha}{\gamma+\beta}
\right)}{(\alpha-\beta)(\alpha+\beta)}.
\end{align}
The above identity can be checked with, {\em e.g.} Ref.~\cite{harris04}.
Finally, after observing that the following identity holds:
\begin{align}
\label{eq20}
G(1,1,1,2,1,1)=-\frac{1}{2w_1}\frac{\partial g}{\partial w_1},
\end{align}
one arrives at the form of the differential equation obeyed by $g$ in the momentum space:
\begin{align}
\label{eq21}
\sigma \frac{\partial g}{\partial w_1}+
\frac{1}{2}\frac{\partial \sigma}{\partial w_1}g(u_1)+
P(w_1,u_1;w_2,u_2;w_3,u_3)=0.
\end{align}
By exchanging the indices at the $\vec{k}$ vectors in the definition of $G(1,1,1,1,1,1)$ one can
obtain analogous differential equations with respect to other variables. In particular, in the
derivation the following one will be required:
\begin{align}
\label{eq22}
\sigma \frac{\partial g}{\partial u_1}+
\frac{1}{2}\frac{\partial \sigma}{\partial u_1}g(u_1)+
P(u_1,w_1;u_3,w_3;w_2,u_2)=0.
\end{align}
The latter two equations were recently presented by Pachucki~\cite{pachucki09}. The solution of this differential
equation was given by Fromm and Hill~\cite{fromm87} and subsequently simplified considerably by
Harris~\cite{harris97}. Unfortunately, the
explicit form of $g$ in terms of well-known special functions is too complicated to perform the inverse
Laplace transform directly and obtain the two-center integrals as in Eq. (\ref{eq3}). Therefore,
the differential equation approach seems to be the only way to derive analytical equations for the
integrals family (\ref{eq0a}).

\subsection{Differential equation in the position space}
\label{subsec:pos}

At this point we will depart from the previous works. To obtain a differential equation for the
master integral $f(r)$ we have to perform the inverse Laplace transform of the Eq. (\ref{eq22}).
Pachucki~\cite{pachucki09} performed such an inversion in the case of $w_1=0$, so any connection with the geminal
basis
was lost. Our case requires a generalization to an arbitrary physically acceptable but nonzero value of $w_1$. Let us
first
rewrite the polynomial $\sigma$ in the following (convenient) way:
\begin{align}
\label{eq23}
\sigma&=w_1^2 u_1^4+\Omega_1 u_1^2+\Omega_2,
\end{align}
\begin{align}
\label{eq23a}
\begin{split}
\Omega_1&=-u_2^2 w_1^2-u_2^2 w_2^2+u_2^2 w_3^2-u_3^2w_1^2+u_3^2 w_2^2\\
&-u_3^2w_3^2+w_1^4-w_1^2w_2^2-w_1^2 w_3^2,
\end{split}
\end{align}
\begin{align}
\begin{split}
\Omega_2&=u_2^4 w_2^2+u_2^2 u_3^2 w_1^2-u_2^2 u_3^2w_2^2-u_2^2 u_3^2 w_3^2\\
&-u_2^2w_1^2 w_2^2+u_2^2w_2^4-u_2^2 w_2^2 w_3^2+u_3^4 w_3^2\\
&-u_3^2 w_1^2w_3^2-u_3^2 w_2^2 w_3^2+u_3^2 w_3^4+w_1^2w_2^2w_3^2,
\end{split}
\end{align}
so that
\begin{align}
\label{eq22a}
\frac{\partial \sigma}{\partial u_1}=4 w_1^2 u_1^3+2 \Omega_1 u_1.
\end{align}
By inserting the above identities into Eq. (\ref{eq22}) and collecting terms multiplying $g(u_1)$ and $\frac{\partial
g}{\partial u_1}$ we get:
\begin{align}
\label{eq23a}
\begin{split}
&\left( w_1^2 u_1^4+\Omega_1 u_1^2+\Omega_2 \right) \frac{\partial g}{\partial u_1}
+\left( 2 w_1^2 u_1^3+\Omega_1 u_1 \right) g(u_1)\\
&+P(u_1,w_1;u_3,w_3;w_2,u_2)=0.
\end{split}
\end{align}
The inverse Laplace transform of this equation leads to:
\begin{align}
\label{eq24}
\begin{split}
&w_1^2 r f^{(4)}(r)+2 w_1^2 f^{(3)}(r)+\Omega_1 r f''(r)\\
&+\Omega_1 f'(r)+\Omega_2 rf(r)=U(r;w_1,u_2,u_3,w_2,w_3),
\end{split}
\end{align}
where
\begin{align}
\label{eq25}
\begin{split}
&U(r;w_1,u_2,u_3,w_2,w_3)=\\
&\frac{1}{2\pi\dot{\imath}}\int_{-\dot{\imath}\infty+\epsilon}^{\dot{\imath}
\infty+\epsilon} du_1\; P(u_1,w_1;u_3,w_3;w_2,u_2) e^{u_1\,r}.
\end{split}
\end{align}
The explicit form of $U(r)$ is obtained by using several Laplace transform identities and reads:
\begin{widetext}
\begin{align}
\label{eq27}
U(r)=\sum_{i=1}^4 c_i U_i(r)+\sum_{i=5}^8 U_i(r),
\end{align}
with
\begin{align}
\label{eq28}
c_1&=\frac{1}{2}\left[w_2 \left(u_2^2-u_3^2-w_1^2\right)+u_3 \left(w_2^2-w_3^2+w_1^2\right)\right],\\
c_2&=\frac{1}{2}\left[w_3 \left(u_2^2-u_3^2+w_1^2\right)+u_2 \left(w_2^2-w_3^2-w_1^2\right)\right],\\
c_3&=\frac{1}{2}\left[w_2 \left(u_2^2-u_3^2-w_1^2\right)-u_3 \left(w_2^2-w_3^2+w_1^2\right)\right],\\
c_4&=\frac{1}{2}\left[u_2 \left(w_2^2-w_3^2-w_1^2\right)-w_3 \left(u_2^2-u_3^2+w_1^2\right)\right],
\end{align}
and
\begin{align}
\label{eq29}
U_1(r)=&e^{r \left(u_3-w_2\right)} \text{Ei}\left[-r
\left(w_1+u_2+u_3\right)\right]-e^{r\left(w_2-u_3\right)}\text{Ei}\left[-r
\left(w_1+w_2+w_3\right)\right],
\end{align}
\begin{align}
U_2(r)=& e^{r \left(w_3-u_2\right)}\text{Ei}\left[-r \left(w_1+w_2+w_3\right)\right]-e^{r
\left(u_2-w_3\right)} \text{Ei}\left[-r\left(w_1+u_2+u_3\right)\right],
\end{align}
\begin{align}
U_3(r)=\notag &e^{-r \left(u_3+w_2\right)}
\left\lbrace\text{Ei}\left[-r\left(w_1+u_2-u_3\right)\right]-\text{Ei}
\left[-r\left(u_2-u_3-w_2+w_3\right)\right]\right.\\
+&\left.\text{Ei}\left[-r \left(w_1-w_2+w_3\right)\right]\right\rbrace
-e^{r \left(u_3+w_2\right)} \text{Ei}\left[-r
\left(u_2+u_3+w_2+w_3\right)\right]\\ \notag
+&e^{-r \left(u_3+w_2\right)}\mbox{Log}\left|\frac{\left(w_1+w_2+w_3\right)
\left(u_2+u_3+w_1\right)
\left(u_2-u_3-w_2+w_3\right)}{\left(w_1-w_2+w_3\right)\left(w_1+u_2-u_3\right)
\left(u_2+u_3+w_2+w_3\right)}\right|,
\end{align}
\begin{align}
U_4(r)=\notag &e^{-r \left(u_2+w_3\right)}
\left\lbrace\text{Ei}\left[-r\left(w_1+u_3-u_2\right)\right]-\text{Ei}
\left[-r\left(u_3-u_2-w_3+w_2\right)\right]\right.\\
+&\left.\text{Ei}\left[-r \left(w_1-w_3+w_2\right)\right]\right\rbrace
-e^{r \left(u_2+w_3\right)} \text{Ei}\left[-r
\left(u_2+u_3+w_2+w_3\right)\right]\\ \notag
+&e^{-r \left(u_2+w_3\right)}\mbox{Log}\left|\frac{\left(w_1+w_2+w_3\right)
\left(u_2+u_3+w_1\right)
\left(u_2-u_3-w_2+w_3\right)}{\left(w_1-w_2+w_3\right)\left(w_1+u_2-u_3\right)
\left(u_2+u_3+w_2+w_3\right)}\right|,
\end{align}
\begin{align}
\label{eq30}
U_5(r)=&-\frac{w_1}{r} e^{-r \left(u_2+w_3\right)} \left(\frac{1}{r}+u_2+w_3\right)+\frac{w_1}{r}
e^{-r \left(u_3+w_1+w_3\right)}\left(\frac{1}{r}+u_3+w_1+w_3\right) \notag \\
-&e^{-r \left(u_2+w_3\right)} w_1
\left[\gamma\left(u_2+w_3\right)\delta(r)+\left(1-\gamma\right)\delta'(r)\right]
+w_1^2 e^{-r\left(u_3+w_1+w_3\right)}\left[\frac{1}{r}+\gamma\,\delta(r) \right] \\ 
+& e^{-r\left(u_3+w_1+w_3\right)}\left[
\gamma\left(u_3+w_3+w_1\right)\delta(r)+\left(1-\gamma\right)\delta'(r) \right] \notag,
\end{align}
\begin{align}
U_6(r)=&-\frac{w_1}{r} e^{-r \left(u_3+w_2\right)} \left(\frac{1}{r}+u_3+w_2\right)+\frac{w_1}{r}
e^{-r \left(u_2+w_1+w_2\right)}\left(\frac{1}{r}+u_2+w_1+w_2\right) \notag \\
-&e^{-r \left(u_2+w_3\right)} w_1
\left[\gamma\left(u_3+w_2\right)\delta(r)+\left(1-\gamma\right)\delta'(r)\right]
+w_1^2 e^{-r\left(u_2+w_1+w_2\right)}\left[\frac{1}{r}+\gamma\,\delta(r) \right]\\ 
+& e^{-r\left(u_2+w_1+w_2\right)}\left[
\gamma\left(u_2+w_2+w_1\right)\delta(r)+\left(1-\gamma\right)\delta'(r) \right] \notag,
\end{align}
\begin{align}
U_7(r)=&w_1^2\,\delta(r)\,\mbox{Log}(w_1+w_2+w_3),\;\;\;
U_8(r)=w_1^2\,\delta(r)\,\mbox{Log}(w_1+u_2+u_3).
\end{align}
\end{widetext}
The results presented above require some comments. First of all let us establish the
connection with the Pachucki differential equation, the zero limit in $w_1$ of Eq.
(\ref{eq24}). By setting $w_1=0$ and observing that:
\begin{align}
\label{eq31}
&p^2=-\frac{\Omega_2}{\Omega_1}\left. \right|_{w_1=0},\;\;\;F(r)=-\frac{U(r;0,u_2,u_3,w_2,w_3)}{\Omega_1},
\end{align}
one arrives at:
\begin{align}
\label{eq32}
r f''(r)+f'(r)-p^2 r f(r)+F(r)=0,
\end{align}
which exactly coincides with the result given by Pachucki~\cite{pachucki09} for the case of orbital basis. Second, at
the end of
this subsection we would like to mention that in the further derivation we will
make use of two other functions which are obtained as the inverse Laplace transforms of $P$, namely:
\begin{align}
\label{eq33}
\begin{split}
&W(r;w_1,u_2,u_3,w_2,w_3)=\\
&\frac{1}{2\pi\dot{\imath}}\int_{-\dot{\imath}\infty+\epsilon}^{\dot{\imath}
\infty+\epsilon} du_1\; P(w_1,u_1;w_2,u_2;w_3,u_3) e^{u_1\,r},
\end{split}
\end{align}
\begin{align}
\label{eq33a}
\begin{split}
&V(r;w_1,u_2,u_3,w_2,w_3)=\\
&\frac{1}{2\pi\dot{\imath}}\int_{-\dot{\imath}\infty+\epsilon}^{\dot{\imath}
\infty+\epsilon} du_1\; P(w_3,u_3;w_2,u_2;w_1,u_1) e^{u_1\,r}
\end{split}
\end{align}
Since explicit formulas for these functions have not been presented in the literature thus far,
we list them in the Appendix \ref{app:appb}.

\section{Solution of the homogeneous differential equation}
\label{sec:hom}

First, we will solve the homogeneous version of the geminal differential equation:
\begin{align}
\label{eq34}
\begin{split}
&w_1^2 r f_H^{(4)}(r)+2 w_1^2 f_H^{(3)}(r)+\Omega_1 r f_H''(r)+\Omega_1 f_H'(r)\\
&+\Omega_2 r f_H(r)=0,
\end{split}
\end{align}
where the subscript $H$ was added to designate the solution of the homogeneous equation. The above equation is a
homogeneous linear ordinary differential equation (ODE) with non-constant coefficients. 
We found it very difficult, if not impossible, to express the solution in terms of well-known special or analytical
functions. Any manipulations performed with Eq. (\ref{eq34}) were proven fruitless in
bringing
this equation into a characteristic form, thus enabling an analytical solution. It was also
impossible to find the solution by using a symbolic mathematical package such as \emph{Mathematica}
\cite{mathematica07}.

It is interesting from the mathematical point of view that Eq. (\ref{eq34}) can be brought
to the form
\begin{align}
\label{eq36}
w_1^2 \frac{d^2}{dr^2}\left(r \frac{d^2f_H}{dr^2} \right)+
\Omega_1 \frac{d}{dr}\left(r \frac{df_H}{dr}\right)=
-r\Omega_2 f_H(r),
\end{align}
so that it can be considered as a generalization of the Sturm-Liouville (S-L) equation to the fourth
order ODE with the weight (or density) function equal to $r$ and eigenvalue $-\Omega_2$.

Because of all the above, we decided to define a new family of special functions which, by
definition, form the general solution of the differetial equation (54). We will find its form by using
the generalized version of the Fr\"obenius method (see, {\em e.g.} Ref.~\cite{smirnoff}). Precisely, we will find a
solution in terms of the series expansion around two singular points, zero and infinity. Our first
Ansatz is an ordinary regular expansion around $r=0$:
\begin{align}
\label{eq37}
f_H(r)=\sum_{k=0}^{\infty} a_k r^k.
\end{align}
We insert this Ansatz into Eq. (\ref{eq34}), collect terms multiplying the
same
power of $r$ and require them to zero to make the differential equation satisfied for all values of
$r$. This establishes the recurrence relation that connects the values of $a_k$ with different $k$.
The final result reads:
\begin{align}
\label{eq38}
\begin{split}
&w_1^2 (k+1)(k+2)^2(k+3)a_{k+3}+\Omega_1(k+1)^2 a_{k+1}+\\
&\Omega_2 a_{k-1}=0\;\;\; \mbox{for}\;\;\; k\geq1,
\end{split}
\end{align}
and the indicial equation:
\begin{align}
\label{eq39}
12w_1^2a_3+\Omega_1a_1=0
\end{align}
Equations (\ref{eq38}) and (\ref{eq39}) need to be simultaneously satisfied. However, there is a freedom
in the choice of three initial
parameters $a_0, a_1$ and $a_2$. Therefore, we specify three new special functions $L_i(r), \;i=1,2,3$, using their
expansions around $r=0$ given by Eq. (\ref{eq37}) and the recurrence relation (\ref{eq38}). The
choice of the three initial parameters is conventional and we put:
\begin{align}
\label{eq40}
\notag L_1(r) \;\;\;\mbox{with}\;\;\; a_0=1, a_1=0, a_2=0, \\
L_2(r) \;\;\;\mbox{with}\;\;\; a_0=0, a_1=1, a_2=0, \\
\notag L_3(r) \;\;\;\mbox{with}\;\;\; a_0=0, a_1=0, a_2=1.
\end{align}
This convention will be used throughout the paper. Let us justify the choice of the formulas (\ref{eq40}). One
may argue that choice
$a_0=1$ in $L_1$ is very special but by putting $a_0=C\neq1$ we obtain a function which is just $L_1$ multiplied by
$C$. The choice of a multiplicative constant is immaterial in our context and, consequently, so is the choice of $C$.
The same is true for the values of $a_1$ and  $a_2$ in $L_2$ and  $L_3$, respectively. Similarly, by defining a function
with $a_0=1,
a_1=1$, for example, we obtain a linear combination of $L_1$ and $L_2$. Because of these properties, we find
the convention (\ref{eq40}) justified.

It is clear that the three functions obtained in the previous paragraph are not sufficient to give the general solution
of
the homogeneous geminal differential equation. Our second trial for the expansion around $r=0$ is somewhat less obvious:
\begin{align}
\label{eq41}
f_H(r)=L_i(r) \mbox{Log}(r)+\sum_{k=0}^\infty b_k r^k,
\end{align}
where the coefficients $b_k$ are to be determined by inserting the expression (\ref{eq41}) into the homogeneous
differential equation and collecting terms
multiplying $r^k$ and $r^k \mbox{Log}(r)$. This results in the recurrence relation:
\begin{align}
\label{eq42}
&2\Omega_1a_1+28w_1^2a_3+\Omega_1b_1+12w_1^2b_3=0,\\
\label{eq42a}
\begin{split}
&2k\Omega_1a_k+2(k+1)(2k^2+4k+1)w_1^2a_{k+2}+\Omega_2b_{k-2}\\
&+k^2\Omega_1b_k+w_1^2k(k+1)^2(k+2)b_{k+2}=0.
\end{split}
\end{align}
Additionally, as soon as $w_1\neq 0$ the above Ansatz requires $a_0=0$, $a_1=1$, $a_2=0$. As
before, we have three parameters which can be chosen freely, $b_0$, $b_1$ and $b_2$. Since we seek
for only one function let us put $b_0=0$, $b_1=1$, $b_2=0$ which leads to:
\begin{align}
\label{eq43}
L_4(r)=L_2(r) \mbox{Log}(r)+\sum_{k=1}^\infty b_k r^k.
\end{align}
One can show that any function constructed with a different choice of $b_0$, $b_1$ and $b_2$
can be expressed as a linear combination of $L_1(r), L_2(r), L_3(r), L_4(r)$. This formally completes
the solution of the homogeneous differential equation (\ref{eq24}).

The expansions around $r=0$ presented above are convergent for all finite values of $r$ since the coefficients
multiplying the powers of $r$ decay faster than any polynomial when $k\rightarrow \infty$. However, the rate of
convergence
of these series can be expected to be poor for large values of $r$ and therefore prohibit an accurate calculation in
this regime. As a result, it might be beneficial to obtain their asymptotic expansion which will be valid and rapidly
convergent for large values of $r$. The latter expansion can be constructed from the Ansatz:
\begin{align}
\label{eq44}
f_H(r)=e^{tr}\sum_{k=0}^\infty a_k r^{-k-\rho},
\end{align}
where $a_k$, $t$ and $\rho$ are coefficients to be determined. By inserting this trial function into the differential
equation
and grouping coefficients multiplying the same powers of $1/r$ one obtains indicial equations
specifying $\rho$ and $t$:
\begin{align}
\label{eq45}
&w_1^2 t^4+\Omega_1 t^2+\Omega_2=0, \\
&\rho=\frac{1}{2},
\end{align}
and the recursion relation for $a_k$ with the value of $\rho$ already fixed at $1/2$:
\begin{align}
\label{eq46}
&\frac{1}{4}\Omega_1 a_0+\frac{3}{2}t^2 w_1^2 a_0-2\Omega_1 t a_1-4t^3 w_1^2 a_1=0, \\
&-3tw_1^2 a_0+\frac{9}{4}\Omega_1 a_1+\frac{27}{2}t^2 w_1^2 a_1-4\Omega_1 t a_2-8t^3w_1^2a_2=0,
\end{align}
and for $k>2$:
\begin{align}
\label{eq47}
\begin{split}
0=&\frac{1}{16}w_1^2 a_k(2k+3)^2(2k+1)(2k+5)\\
-&tw_1^2 a_{k+1}(k+2)(2k+3)(2k+5)\\
+&\frac{1}{4}(2k+5)^2a_{k+2}\left(\Omega_1+6t^2w_1^2\right)\\
-&2(k+3)a_{k+3}\left(\Omega_1t+2t^3w_1^2\right).
\end{split}
\end{align}
Eq. (\ref{eq45}) has four solutions $t_i$, $i=1,...,4$, which correspond to four functions
determining the general
solution of the homogeneous differential equation. We see that it is dependent on the sign of $t_i$ whether convergent
or divergent expansion is obtained. The final result can be written as:
\begin{align}
\label{eq48}
f_H(r)=\frac{e^{t_i r}}{\sqrt{r}}\sum_{k=0}^\infty \frac{a_k}{r^k}
\end{align}
where $a_0$ can freely be chosen.

It is interesting to establish a connection between the new special functions $L_i$ and
the modified
Bessel functions of the first, $I_0(r)$, and the second, $K_0(r)$, kind. By setting $w_1=0$
Eqs. (\ref{eq38}) and (\ref{eq39}) become:
\begin{align}
\label{eq49}
\Omega_1(k+1)^2 a_{k+1}+\Omega_2 a_{k-1}=0, \;\;\;
a_1=0,
\end{align}
so that the recursion can be solved explicitly to give:
\begin{align}
\label{eq50}
a_{2k}=\frac{\left.-\frac{\Omega_2}{\Omega_1}\right|_{w_1=0}^k}{2^{2k} k!^2 }=\frac{p^{2k}}{2^{2k}
k!^2},\;\;\;
a_{2k-1}=0,
\end{align}
and the series can be brought into the closed form:
\begin{align}
\label{eq51}
\sum_{k=0}^\infty \frac{p^{2k}r^{2k}}{2^{2k} k!^2}=I_0(pr),
\end{align}
coinciding with the Bessel function of the first kind. Similarly, by setting $w_1=0$ in Eqs. (\ref{eq42}) and
(\ref{eq42a}) one finds a linear combination of $I_0(pr)$ and
$K_0(pr)$ to be the $w_1=0$ limit of $L_4(r)$. One could force the exact relationship:
\begin{align}
\label{eq52}
\lim_{w_1\rightarrow 0} L_4(r)=K_0(pr),
\end{align}
by a proper choice of the initial parameters. Our choice was made for the sake of
simplicity as indicated before. Similar result is found with the asymptotic expansions of $L_i(r)$.
Whenever $w_1=0$, Eq. (\ref{eq45}) has two solutions:
\begin{align}
\label{eq53}
t_\pm=\pm \sqrt{-\frac{\Omega_2}{\Omega_1}}=\pm p,
\end{align}
so that Eq. (\ref{eq48}) becomes the asymptotic expansion of $I_0$ (with $t=p$) or $K_0$ (with
$t=-p$).

We believe that because of the interesting properties of the $L_i(r)$ functions and their strong connection with the
Bessel functions they can be understood as a generalization to the fourth order differential
equation. Therefore, we give them the name \emph{hyper-Bessel functions}. In analogy, $L_1,
L_2, L_3$ functions are hyper-Bessel functions of the first kind and $L_4$ is the hyper-Bessel
function of the second kind. 

\section{Solution of the inhomogeneous differential equation}
\label{sec:inhom}

The next step in our derivation is to use the properties of the functions introduced in Sec. \ref{sec:hom} to obtain
solution of the inhomogeneous differential equation (\ref{eq24}). 
In this work we decided to use the method based on the Wronskian determinants. Starting with the general solution of the
homogeneous
equation:
\begin{align}
\label{eq54}
f_H(r)=c_1 L_1(r)+c_2 L_2(r)+c_3 L_3(r)+c_4 L_4(r),
\end{align}
we make coefficients $c_i$ explicit functions of $r$, $y_i(r)$, and require the combination
\begin{align}
\label{eq55}
\begin{split}
f(r)&=y_1(r) L_1(r)+y_2(r) L_2(r)\\
&+y_3(r) L_3(r)+y_4(r) L_4(r),
\end{split}
\end{align}
to satisfy the inhomogeneous geminal equation (\ref{eq24}). Differentiation of the above equation leads to:
\begin{align}
\label{eq56}
\begin{split}
f'(r)=&\left[ y_1(r) L'_1(r)+y_2(r) L'_2(r)\right.\\
+&\left.y_3(r) L'_3(r)+y_4(r) L'_4(r) \right]\\
+&\left[ y'_1(r) L_1(r)+y'_2(r) L_2(r)\right.\\
+&\left.y'_3(r) L_3(r)+y'_4(r) L_4(r) \right].
\end{split}
\end{align}
Requirement that the second expression vanishes identically for all $r$:
\begin{align}
\label{eq57}
\begin{split}
0&\equiv y'_1(r) L_1(r)+y'_2(r) L_2(r)\\
&+y'_3(r) L_3(r)+y'_4(r) L_4(r),
\end{split}
\end{align}
enables us bring the first derivative into the form:
\begin{align}
\label{eq58}
\begin{split}
f'(r)&= y_1(r) L'_1(r)+y_2(r) L'_2(r)\\
&+y_3(r) L'_3(r)+y_4(r) L'_4(r).
\end{split}
\end{align}
Similarly, higher-order derivatives are found to be:
\begin{align}
\label{eq59}
\begin{split}
f''(r)&=y_1(r) L''_1(r)+y_2(r) L''_2(r)\\
&+y_3(r) L''_3(r)+y_4(r) L''_4(r),
\end{split}
\end{align}
\begin{align}
\label{eq59a}
\begin{split}
f^{(3)}(r)&=y_1(r) L^{(3)}_1(r)+y_2(r) L^{(3)}_2(r)\\
&+y_3(r) L^{(3)}_3(r)+y_4(r) L^{(3)}_4(r),
\end{split}
\end{align}
\begin{align}
\label{eq59b}
\begin{split}
f^{(4)}(r)&=\left[y_1(r) L^{(4)}_1(r)+y_2(r) L^{(4)}_2(r)\right.\\
&\left.+y_3(r) L^{(4)}_2(r)+y_4(r)
L^{(4)}_4(r)\right]\\
&+\left[ y'_1(r) L^{(3)}_1(r)+y'_2(r) L^{(3)}_2(r)\right.\\
&+\left.y'_3(r) L^{(3)}_3(r)+y'_4(r) L^{(3)}_4(r) \right],
\end{split}
\end{align}
where additional constraints on $y_i(r)$ where imposed:
\begin{align}
\label{eq60a}
&y'_1(r) L'_1(r)+y'_2(r) L'_2(r)+y'_3(r) L'_3(r)+y'_4(r) L'_4(r)\equiv0, \\
\label{eq60b}
&y'_1(r) L''_1(r)+y'_2(r) L''_2(r)+y'_3(r) L''_3(r)+y'_4(r) L''_4(r)\equiv0.
\end{align}
By inserting Eqs. (\ref{eq59}), (\ref{eq59a}), and (\ref{eq59b}) into the differential equation and noting that
the functions $L_i(r)$ satisfy the homogeneous differential equation one arrives at:
\begin{align}
\label{eq61}
U(r)=&w_1^2 r \left[y'_1(r) L^{(3)}_1(r)+y'_2(r) L^{(3)}_2(r)\right.\\
+&\left.y'_3(r) L^{(3)}_3(r)+y'_4(r) L^{(3)}_4(r)\right].
\end{align}
The above equation together with Eqs. (\ref{eq57}), (\ref{eq60a}), and (\ref{eq60b}) specify the
four-dimensional system of linear equations:
\begin{align}
\label{eq62}
\left[ \begin{array}{cccc}
L^{(3)}_1(r) & L^{(3)}_2(r) & L^{(3)}_3(r) & L^{(3)}_4(r) \\
L''_1(r) & L''_2(r) & L''_3(r) & L''_4(r) \\
L'_1(r) & L'_2(r) & L'_3(r) & L'_4(r) \\
L_1(r) & L_2(r) & L_3(r) & L_4(r)
\end{array} \right]
\left[ \begin{array}{c}
y'_1(r) \\
y'_2(r) \\
y'_3(r) \\
y'_4(r)
\end{array} \right]=
\left[ \begin{array}{c}
\frac{U(r)}{w_1^2 r}\\
0 \\
0 \\
0
\end{array} \right],
\end{align}
that can easily be solved symbolically for $y'_i(r)$ by using the Cramer's rule. The result reads:
\begin{align}
\label{eq63}
y'_1(r)=\frac{U(r)}{w_1^2 r} \frac{W_1(r)}{W(r)},
\end{align}
where $W(r)$ is the Wronskian determinant of the functions $L_1(r), L_2(r), L_3(r), L_4(r)$:
\begin{align}
\label{eq64}
W(r)=\left| \begin{array}{cccc}
L^{(3)}_1(r) & L^{(3)}_2(r) & L^{(3)}_3(r) & L^{(3)}_4(r) \\
L''_1(r) & L''_2(r) & L''_3(r) & L''_4(r) \\
L'_1(r) & L'_2(r) & L'_3(r) & L'_4(r) \\
L_1(r) & L_2(r) & L_3(r) & L_4(r)
\end{array} \right|,
\end{align}
and $W_k(r)$ are the same as $W(r)$ apart form the fact that the$k$th column was replaced by the unit vector
$[1,0,0,0]$,
for example:
\begin{align}
\label{eq65}
\begin{split}
W_1(r)&=\left| \begin{array}{cccc}
1 & L^{(3)}_2(r) & L^{(3)}_3(r) & L^{(3)}_4(r) \\
0 & L''_2(r) & L''_3(r) & L''_4(r) \\
0 & L'_2(r) & L'_3(r) & L'_4(r) \\
0 & L_2(r) & L_3(r) & L_4(r)
\end{array} \right|\\
&=\left| \begin{array}{ccc}
L''_2(r) & L''_3(r) & L''_4(r) \\
L'_2(r) & L'_3(r) & L'_4(r) \\
L_2(r) & L_3(r) & L_4(r)
\end{array} \right|.
\end{split}
\end{align}
Some degree of suspicion may be connected with the fact that Wronskian appears in the denominator. However, since the
functions $L_1(r), L_2(r), L_3(r)$, and $L_4(r)$ span the space of solutions of the homogeneous differential equation
they
cannot be linearly dependent and thus $W(r)$ cannot vanish identically. Eq. (\ref{eq63}) can now formally be
integrated
\begin{align}
\label{eq66}
y_1(r)=\int dr \frac{U(r)}{w_1^2 r} \frac{W_1(r)}{W(r)},
\end{align}
so that the solution of the inhomogeneous differential equation is
\begin{align}
\label{eq67}
f(r)&=\sum_{i=1}^4
L_i(r)\int dr \frac{U(r)}{w_1^2 r} \frac{W_i(r)}{W(r)},
\end{align}
where the initial conditions have not been imposed yet. At this point we observe that the solution is rather
complicated because of the presence of five determinants, including the Wronskian itself which is the most cumbersome in
the calculations. Therefore, it will be advantageous to introduce some simplifications in the above formula. It
turns out that the appearance of the Wronskian can be eliminated altogether by using the so-called Abel's identity which
states
that for any $n$th-order homogeneous differential equation of the form:
\begin{align}
\label{eq68}
y^{(n)}(x)+p_{n-1}(x)y^{(n-1)}(x)+\ldots+p_0(x)y(x)=0,
\end{align}
the Wronskian constructed from $n$ linearly independent solutions can be expressed as:
\begin{align}
\label{eq69}
W(x)=W(x_0)\exp\left(- \int_{x_0}^x dx' p_{n-1}(x') \right),
\end{align}
provided that $p_{n-1}(x)$ is continuous on the interval $\left[ x_0, x\right]$. In the case of the geminal differential
equation
$p_{n-1}(r)=\frac{2}{r}$ (continuous on $0< r\leq \infty$), so that integration can easily be carried out and the Abel's
identity is:
\begin{align}
\label{eq70}
W(r)=W(r_0)\left( \frac{r_0}{r} \right)^2.
\end{align}
We need to specify the point $r_0$. Our choice $r_0=1$ was motivated by the fact that the above
equation takes a very simple form and that $r_0=1$ is sufficiently close to $r=0$ at which series expansions of
$L_i(r)$
were provided. It allows a robust calculation of $W(1)$ for any values of the nonlinear parameters.
By inserting the identity:
\begin{align}
\label{eq71}
W(r)=W(1) \frac{1}{r^2}, 
\end{align}
into the solution (\ref{eq67}) considerable simplifications occur:
\begin{align}
\label{eq72}
f(r)=\sum_{i=1}^4
\frac{L_i(r)}{w_1^2 W(1)}\int dr\; r\, U(r) W_i(r).
\end{align}
Despite a considerable effort we did not manage to simplify this equation further, at least in the general case. Such a
simplification will occur for a special case considered in the next subsection.

Finally, we have to impose four initial conditions on the above solution to make it consistent with the definition of
the master integral. The first three initial conditions are natural:
\begin{align}
\label{eq73a}
f(0)=0,\;\;\;
\lim_{r\rightarrow \infty} f(r)=0,\;\;\;
\lim_{r\rightarrow \infty} f'(r)=0.
\end{align}
The fourth initial condition is somehow more complicated, but we see that whenever $r \rightarrow 0$ then
$r_{1B}\rightarrow r_{1A}=r_1$ etc., so that in the $r=0$ limit the derivative of the master integral becomes the
well-known integral:
\begin{equation}
\label{eq74}
f'(0)=\int \frac{d^3 r_1}{4\pi}\int \frac{d^3 r_2}{4\pi}
\frac{e^{-(u_3+u_2) \,r_{1}}}{r_{1}^2} \frac{e^{-(w_2+w_3) \,r_{2}}}{r_{2}^2} \frac{e^{-w_1 \,r_{12}}}{r_{12}}.
\end{equation}
Analytical formula for this integral is well-known, \emph{cf.} Eq. (34) of Ref.~\cite{harris04}.
After the four initial conditions are imposed the master integral becomes:
\begin{widetext}
\begin{align}
\label{eq76}
\begin{split}
f(r)&=
\frac{L_1(r)}{w_1^2 W(1)}\int_0^r dr'\; r'\, U(r') W_1(r') 
-\frac{L_2(r)}{w_1^2 W(1)}\int_r^\infty dr'\; r'\, U(r') W_2(r') \\&-
\frac{L_3(r)}{w_1^2 W(1)}\int_r^\infty\ dr'\; r'\, U(r') W_3(r') +
\frac{L_4(r)}{w_1^2 W(1)}\int_0^r dr'\; r'\, U(r') W_4(r'). 
\end{split}
\end{align}
\end{widetext}
One can check that the above formula satisfies the initial conditions
(\ref{eq73a}) and (\ref{eq74}), and therefore is the solution of the inhomogeneous differential geminal
equation. In the further derivation
we will also need the values of $f'(r)$, $f''(r)$ and $f^{(3)}(r)$. Higher-order derivatives can be obtained
recursively by differentiation of Eq. (\ref{eq24}):
\begin{align}
\label{eq77}
\begin{split}
f^{(n+4)}(r)&=\frac{U^{(n)}(r)}{rw_1^2}
-\frac{n+2}{r}f^{(n+3)}(r)-\frac{\Omega_1}{w_1^2}f^{(n+2)}(r)\\
&-\frac{\Omega_1(n+1)}{rw_1^2}f^{(n+1)}(r)
-\frac{\Omega_2}{w_1^2}f^{(n)}-\frac{\Omega_2 n}{rw_1^2}f^{(n-1)}.
\end{split}
\end{align}
\begin{widetext}
The first derivative of the master integral with respect to $r$ is obtained directly from the
representation (\ref{eq76}):
\begin{align}
\begin{split}
f'(r)&=
\frac{L'_1(r)}{w_1^2 W(1)}\int_0^r dr'\; r'\, U(r') W_1(r')
-\frac{L'_2(r)}{w_1^2 W(1)}\int_r^\infty dr'\; r'\, U(r') W_2(r') \\
&-\frac{L'_3(r)}{w_1^2 W(1)}\int_0^\infty\ dr'\; r'\, U(r') W_3(r')
+\frac{L'_4(r)}{w_1^2 W(1)}\int_0^r dr'\; r'\, U(r') W_4(r') \\
&+\sum_{i=1}^4 \frac{L_i(r)}{w_1^2 W(1)} r\, U(r) W_i(r).
\end{split}
\end{align}
\end{widetext}
The non-integral term is equal to:
\begin{align}
&\sum_{i=1}^4 \frac{L_i(r)}{w_1^2 W(1)} r\, U(r) W_i(r)= \sum_{i=1}^4 L_i(r) y'_i(r),
\end{align}
so it vanishes identically on the basis of the initial assumption (\ref{eq57}). The first derivative of
$f(r)$ becomes:
\begin{widetext}
\begin{align}
\label{eq77a}
\begin{split}
f'(r)&=
\frac{L'_1(r)}{w_1^2 W(1)}\int_0^r dr'\; r'\, U(r') W_1(r') 
-\frac{L'_2(r)}{w_1^2 W(1)}\int_r^\infty dr'\; r'\, U(r') W_2(r') \\&-
\frac{L'_3(r)}{w_1^2 W(1)}\int_r^\infty\ dr'\; r'\, U(r') W_3(r') +
\frac{L'_4(r)}{w_1^2 W(1)}\int_0^r dr'\; r'\, U(r') W_4(r').
\end{split}
\end{align}
Similarly, using the conditions (\ref{eq60a}) and (\ref{eq60b}), explicit formulas for $f''(r)$ and
$f^{(3)}(r)$ are obtained:
\begin{align}
\label{eq77b}
\begin{split}
f''(r)&=
\frac{L''_1(r)}{w_1^2 W(1)}\int_0^r dr'\; r'\, U(r') W_1(r') 
-\frac{L''_2(r)}{w_1^2 W(1)}\int_r^\infty dr'\; r'\, U(r') W_2(r') \\&-
\frac{L''_3(r)}{w_1^2 W(1)}\int_r^\infty\ dr'\; r'\, U(r') W_3(r') +
\frac{L''_4(r)}{w_1^2 W(1)}\int_0^r dr'\; r'\, U(r') W_4(r'),
\end{split}
\end{align}
\begin{align}
\label{eq77c}
\begin{split}
f^{(3)}(r)&=
\frac{L^{(3)}_1(r)}{w_1^2 W(1)}\int_0^r dr'\; r'\, U(r') W_1(r') 
-\frac{L^{(3)}_2(r)}{w_1^2 W(1)}\int_r^\infty dr'\; r'\, U(r') W_2(r') \\&-
\frac{L^{(3)}_3(r)}{w_1^2 W(1)}\int_r^\infty\ dr'\; r'\, U(r') W_3(r') +
\frac{L^{(3)}_4(r)}{w_1^2 W(1)}\int_0^r dr'\; r'\, U(r') W_4(r'),
\end{split}
\end{align}
\end{widetext}
so that values of the latter three quantities can be calculated with an insignificant additional cost once the numerical
integration of the integrals appearing in $f(r)$ is done.


\section{Recursion relations for the powers of $r_{1A}, r_{1B}, r_{2A}, r_{2B}$}
\label{sec:recr1a}

With the value of the master integral at hand, we turn to the calculation of the integrals with
arbitrary powers of $r_{1A}, r_{1B}, r_{2A}, r_{2B}$. They are obtained by differentiation of the
master integral with respect to the nonlinear parameters. Explicit differentiation of Eq.
(104) is cumbersome and connected with painful and expensive numerical
integrations. Therefore, to start the recursion relations, we must establish an equation that
connects the value of the derivative of the master integral with respect to, say, $w_3$, to the
master integral and optionally its derivatives with respect to $r$. The latter quantities can be
computed by using the theory presented in the previous section.

The desired recursion relation can be derived from two differential equations in the momentum
space. The first was already derived in the subsection II B:
\begin{align}
\label{eq83}
\sigma \frac{\partial g}{\partial u_1}+
\frac{1}{2}\frac{\partial \sigma}{\partial u_1}g(u_1)+
P(u_1,w_1;u_3,w_3;w_2,u_2)=0,
\end{align}
and the second is obtained by the proper exchange of the nonlinear parameters, making use of the
fact that both $g$ and $\sigma$ are invariant under the latter operations:
\begin{align}
\label{eq84}
\sigma \frac{\partial g}{\partial w_3}+
\frac{1}{2}\frac{\partial \sigma}{\partial w_3}g(u_1)+
P(w_3,u_3;w_2,u_2;w_1,u_1)=0.
\end{align}
By taking the inverse Laplace transform of both equations one obtains
\begin{align}
\begin{split}
-w_1^2 r f^{(4)}(r)-2 w_1^2 f^{(3)}(r)-\Omega_1 r f''(r)-\Omega_1 f'(r)\\
-\Omega_2 rf(r)+U(r;w_1,u_2,u_3,w_2,w_3)=0,
\end{split}
\end{align}
\begin{align}
\begin{split}
w_1^2 \frac{\partial f^{(4)}}{\partial w_3}+
\Omega_1 \frac{\partial f''}{\partial w_3}+
\Omega_2 \frac{\partial f}{\partial w_3}+
\frac{1}{2}\frac{\partial \Omega_1}{\partial w_3}f''(r)\\
+\frac{1}{2}\frac{\partial \Omega_2}{\partial w_3}f(r)
+V(r;w_1,u_2,u_3,w_2,w_3)=0,
\end{split}
\end{align}
and by differentiation of the first equation with respect to $w_3$ one obtains a pair:
\begin{widetext}
\begin{align}
\label{eq86}
\notag \mbox{E}_1\equiv&-w_1^2\, r\, \frac{\partial f^{(4)}}{\partial w_3}-2 w_1^2 \,\frac{\partial
f^{(3)}}{\partial w_3}(r)
-\frac{\partial \Omega_1}{\partial w_3} r f''(r)
-\Omega_1 r \frac{\partial f''}{\partial w_3}
-\frac{\partial \Omega_1}{\partial w_3} f'(r)
-\Omega_1 \frac{\partial f'}{\partial w_3}\\
&-\frac{\partial \Omega_2}{\partial w_3} rf(r)
-\Omega_2 r \frac{\partial f}{\partial w_3}
+\frac{\partial U(r)}{\partial w_3}=0, \\
\notag \mbox{E}_2\equiv&w_1^2\, \frac{\partial f^{(4)}}{\partial w_3}+
\Omega_1 \frac{\partial f''}{\partial w_3}+
\Omega_2 \frac{\partial f}{\partial w_3}+
\frac{1}{2}\frac{\partial \Omega_1}{\partial w_3}f''(r)+
\frac{1}{2}\frac{\partial \Omega_2}{\partial w_3}f(r)+V(r)=0,
\end{align}
\end{widetext}
where the notation for the nonlinear parameters in $U$ and $V$ was suppressed for brevity. These
two equations provide a starting point to establish an explicit recursion relation. However, its
derivation is still a nontrivial task since $\mbox{E}_1$, $\mbox{E}_2$, apart from the desired term
$\frac{\partial f}{\partial w_3}$, consist of the derivatives of the latter with respect to $r$ up to the
fourth order. Our approach was based on the following three additional identities that are defined as:
\begin{align}
\label{eq88}
\mbox{E}_3&=\frac{\partial}{\partial r}\left(\mbox{E}_1+r\, \mbox{E}_2\right), \\
\mbox{E}_4&=\frac{\partial}{\partial r}\left(r\, \mbox{E}_3-2\,\mbox{E}_1\right), \\
\mbox{E}_5&=\frac{\partial}{\partial r}\left(\mbox{E}_4-4\,\mbox{E}_2 \right).
\end{align}
The reason for making the combinations above is as follows. At each step we cancel out the
fourth-order derivative of $\frac{\partial f}{\partial w_3}$ with respect to $r$ and then create it
back by doing a differentiation with respect to $r$. By repeating this procedure three times we
figure out that the Eq. (\ref{eq88}) is a set of equations with five unknown quantities:
\begin{align}
\notag 
\frac{\partial f}{\partial w_3},\;
\frac{\partial f'}{\partial w_3},\;
\frac{\partial f''}{\partial w_3},\;
\frac{\partial f^{(3)}}{\partial w_3},\;
\frac{\partial f^{(4)}}{\partial w_3},
\end{align}
so it can be solved analytically.
The differentiation performed at each step guarantees that $\mbox{E}_{i}$, $i=1,5$, are linearly
independent
as long as none of the coefficients multiplying the unknown quantities in the initial equations
for $\mbox{E}_1$ and $\mbox{E}_2$ vanishes. Higher-order derivatives of $\frac{\partial f}{\partial w_3}$
over $r$ do not appear. The final result is:
\begin{align}
\label{eq90}
\begin{split}
 &\frac{\partial f}{\partial w_3}=\\
 &2w_1^2 \frac{\Omega_2}{D_0} \left\lbrace
6V(r)+2rV'(r)+2\frac{\partial U'(r)}{\partial w_3}
+\frac{\partial \Omega_2}{\partial w_3}[f(r)+rf'(r)]\right.\\
&-\left.\frac{\partial \Omega_1}{\partial w_3}[f''(r)+rf^{(3)}(r)]
\right\rbrace 
+\frac{\Omega_1^2}{D_0}\left\lbrace 
-4 V(r)-2r V'(r)\right.\\
&\left.-2\frac{\partial U'(r)}{\partial w_3}+
\frac{\partial \Omega_2}{\partial w_3}r f'(r)+
\frac{\partial \Omega_1}{\partial w_3} [2f''(r)+rf^{(3)}(r)]
\right\rbrace \\
&+ w_1^2\frac{ \Omega_1}{D_0}\left\lbrace
-10 V''(r)-2rV^{(3)}(r)-2\frac{\partial U^{(3)}(r)}{\partial w_3}\right.\\
&+\left.\frac{\partial \Omega_2}{\partial w_3} [f''(r)+rf^{(3)}(r)]+
\frac{\partial \Omega_1}{\partial w_3} [3f^{(4)}(r)+rf^{(5)}(r)]
\right\rbrace,
\end{split}
\end{align}
where $D_0$ is the common denominator:
\begin{align}
\label{eq91}
D_0=2\Omega_2\left(\Omega_1^2-4w_1^2\Omega_2\right).
\end{align}
It is noteworthy that the procedure in which the required set $\mbox{E}_{i}$, $i=1,5$, was obtained is
somehow ambiguous. Only the first step of this procedure, formation of $\mbox{E}_3$, is unique
since there is only one correct method to obtain a useful equation by cancelling out the
fourth-order derivative of $\frac{\partial f}{\partial w_3}$. In the further steps such an
elimination can be performed using different equations which were obtained previously and  the number of
possibilities grows with the number of steps taken. We cannot prove that the particular choice of
equations for $\mbox{E}_{i}$, $i=1,5$, which we used here is `the best'. However, in our procedure we
tried to minimize the order of the derivatives of functions $U(r)$ and $V(r)$ that appeared in the
final result. It leads to equation (\ref{eq90}) which turned out to  regular.

By multiplying both sides of the relation (\ref{eq90}) by $D_0$ and by further
differentiation one
can calculate arbitrary derivative over the nonlinear parameters thus advancing the powers of
$r_{1A}, r_{1B}, r_{2A}, r_{2B}$ as much as necessary. The recursion relations for the derivatives over
$w_2$, $u_3$ and $u_2$ that cannot directly be calculated from the above formula are obtained with
the use of the symmetry of the master integral. Namely, by permuting $w_2\leftrightarrow w_3$ and
$u_2\leftrightarrow u_3$ (exchange of the nuclei $A\leftrightarrow B$ in the master integral) and noting that the 
master integral is invariant with respect to this permutation, analogous
recursion
relation for $\frac{\partial f}{\partial w_2}$ is obtained. Similarly, the exchange of
$u_2\leftrightarrow w_3$ and $u_3\leftrightarrow w_2$ (change of the electrons' numbering
$1\leftrightarrow 2$) results in the derivative over $u_2$. Finally, the use of both of these 
permutations gives the derivative over $u_3$.

We listed only the formula for $\frac{\partial f}{\partial w_3}$ despite the fact that
by solving the set of equations for $\mbox{E}_{i}$, $i=1,5$, its derivatives over $r$ up to the fourth-order
are obtained as by-products. From the mathematical point of view equivalent formulas can be
derived by differentiating Eq. (\ref{eq90}) over $r$. Although numerical results obtained in this manner are
the same, formulas for higher-order derivatives over $r$ calculated from the solution of
Eqs. (\ref{eq88}) are much more transparent. In particular, they do not include higher-order
derivatives of the functions $U(r)$, $V(r)$, and of the master integral. Therefore, we list all the
missing formulas in the Appendix \ref{app:appc}.

\section{Recursion relations for the powers of $r_{12}$}
\label{sec:recr12}

Since the integrals in the Slater geminal basis considered here already include explicit correlation
factor, there is a little point in growing powers of $r_{12}$ in the initial basis set.
However, such a possibility is open and we will elaborate it in this
section. In particular, we will derive an analytical equation for the overlap integral over
Slater geminals which, despite its simplicity at first glance, has not found analytical
solution yet. Our approach is similar to the one presented in the previous section. We shall
establish a relation that connects the value of $\frac{\partial f}{\partial w_1}$ with the master integral and its
derivatives over $r$. In the
derivation we use the following differential equations for $g$ in the momentum space:
\begin{align}
\label{eq92}
\sigma \frac{\partial g}{\partial u_1}+
\frac{1}{2}\frac{\partial \sigma}{\partial u_1}g(u_1)+
P(u_1,w_1;u_3,w_3;w_2,u_2)=0, \\
\sigma \frac{\partial g}{\partial w_1}+
\frac{1}{2}\frac{\partial \sigma}{\partial w_1}g(u_1)+
P(w_1,u_1;w_2,u_2;w_3,u_3)=0.
\end{align}
The first of these equations is differentiated with respect to $w_1$ and then the inverse Laplace transform is performed
to give:
\begin{widetext}
\begin{align}
\label{eq93}
\begin{split}
\overline{\mbox{E}}_1&=-rw_1^2 \frac{\partial f^{(4)}}{\partial w_1}-2w_1^2\frac{\partial f^{(3)}}{\partial w_1}
-\Omega_1 r \frac{\partial f''}{\partial w_1}-\Omega_1 \frac{\partial f'}{\partial w_1}
-\Omega_2 r \frac{\partial f}{\partial w_1}\\
 &-2rw_1 f^{(4)}(r)-4w_1f^{(3)}(r)-r\frac{\partial \Omega_1}{\partial w_1}f''(r)
-\frac{\partial \Omega_1}{\partial w_1}f'(r)-r\frac{\partial \Omega_2}{\partial w_1}f(r)+
\frac{\partial U(r)}{\partial w_1}=0,\\
\end{split}
\end{align}
\begin{align}
\label{eq93a}
\overline{\mbox{E}}_2&=w_1^2 \frac{\partial f^{(4)}}{\partial w_1}+\Omega_1 \frac{\partial f''}{\partial w_1}
+\Omega_2 \frac{\partial f}{\partial w_1}+w_1 f^{(4)}(r)+\frac{1}{2}\frac{\partial \Omega_1}{\partial w_1} f''(r)+
\frac{1}{2}\frac{\partial \Omega_2}{\partial w_1}f(r)+W(r)=0,
\end{align}
\end{widetext}
Using a similar procedure as for the derivatives over $w_3$ we form a set of equations:
\begin{align}
\label{eq94}
&\overline{\mbox{E}}_3=\frac{\partial}{\partial r}\left( \overline{\mbox{E}}_1+r\overline{\mbox{E}}_2 \right),\\
&\overline{\mbox{E}}_4=\frac{\partial}{\partial r}\left( \overline{\mbox{E}}_3+2\overline{\mbox{E}}_2 \right),\\
&\overline{\mbox{E}}_5=\frac{\partial \overline{\mbox{E}}_4}{\partial r},
\end{align}
which are then solved for the following quantities:
\begin{align}
\notag 
\frac{\partial f}{\partial w_1},\;
\frac{\partial f'}{\partial w_1},\;
\frac{\partial f''}{\partial w_1},\;
\frac{\partial f^{(3)}}{\partial w_1},\;
\frac{\partial f^{(4)}}{\partial w_1}.
\end{align}
The final equation for $\frac{\partial f}{\partial w_1}$ is given by:
\begin{widetext}
\begin{align}
\label{eq96}
\begin{split}
 \frac{\partial f}{\partial w_1}=&\\
 -&2w_1^2 \frac{\Omega_2}{D_0} \left\lbrace
-6W(r)-2rW'(r)-2\frac{\partial U'(r)}{\partial w_1}
-\frac{\partial \Omega_2}{\partial w_1}[f(r)-rf'(r)]\right. \\
 +&\left.\frac{\partial \Omega_1}{\partial w_1}[f''(r)+rf^{(3)}(r)]+
6w_1f^{(4)}(r)+2w_1f^{(5)}(r)
\right\rbrace \\
 +&\frac{\Omega_1^2}{D_0}\left\lbrace 
-4 W(r)-2r W'(r)-2\frac{\partial U'(r)}{\partial w_1}+
\frac{\partial \Omega_2}{\partial w_1}r f'(r)+
\frac{\partial \Omega_1}{\partial w_1} [2f''(r)+rf^{(3)}(r)]\right.+\\
+&\left. 8w_1f^{(4)}(r)+2rw_1f^{(5)}(r)
\right\rbrace \\
 +& w_1^2\frac{\Omega_1}{D_0}\left\lbrace
-10 W''(r)-2rW^{(3)}(r)-2\frac{\partial U^{(3)}(r)}{\partial w_1}+
\frac{\partial \Omega_2}{\partial w_3} [f''(r)+rf^{(3)}(r)]\right.\\
 +&\left.\frac{\partial \Omega_1}{\partial w_3} [3f^{(4)}(r)+rf^{(5)}(r)]
+10w_1 f^{(6)}(r)+2w_1 r f^{(7)}(r)
\right\rbrace.
\end{split}
\end{align}
\end{widetext}
Higher powers of $r_{12}$ are obtained by further differentiation of the above equation. As before, useful
formulas resulting from the solution of the set for $\overline{\mbox{E}}_{i}$,
$i=1,5$, are listed in the Appendix \ref{app:appc}.

\section{Special cases}
\label{sec:special}

In this section we consider four special cases of the integrals corresponding to
situations when coefficients $\Omega_1$ and/or $\Omega_2$ vanish. From the mathematical point of view
the recursion relations
for the coefficients $a_k$ and $b_k$ remain valid since in their recursive evaluation one never divides
by $\Omega_1$ or $\Omega_2$. Therefore, the representation of the master integral given by Eq. (\ref{eq74}) is still
correct. However, for practical reasons it is useful to consider these two special cases in details
since the solution of the homogeneous differential equation can be expressed in terms of well-known special functions.
This
makes the implementation of the method much simpler.

\subsection{Vanishing $\Omega_2$ coefficient}
\label{subsec:omega2}

Vanishing $\Omega_2$ coefficient is probably the most important special case since it occurs for a
handful of physically important classes of integrals. This includes exponentially correlated
analogue of the symmetric James-Coolidge basis set~\cite{james33} ($u_2=u_3=w_2=w_3=x$) and symmetric exchange
integrals over
atomic orbitals ($u_3=w_2=x$, $u_2=w_3=y$). Singularities also appear whenever:
\begin{align}
w_1^2=\left(u_2^2w_2^2-u_3^2w_3^2\right)\left(\frac{1}{u_2^2-w_3^2}-\frac{1}{u_3^2-w_2^2}\right).
\end{align}
Moreover, the recursion relations established in the previous subsections are not valid in
this case since $\Omega_2$ appears in the denominator in the key formulas.

In this special case the geminal differential equation takes the form:
\begin{align}
\label{eq97}
w_1^2 r f^{(4)}(r)+2 w_1^2 f^{(3)}(r)+\Omega_1 r f''(r)+\Omega_1 f'(r)=U(r),
\end{align}
so that the homogeneous equation is:
\begin{align}
\label{eq98}
w_1^2 r f_H^{(4)}(r)+2 w_1^2 f_H^{(3)}(r)+\Omega_1 r f_H''(r)+\Omega_1 f_H'(r)=0.
\end{align}
The simplest way to obtain the solution of the latter equation is to use the recursion relations for the
coefficients in the series expansions that were derived for the general case, Eqs. (\ref{eq38}), (\ref{eq39}),
(\ref{eq42}), and (\ref{eq42a}). By setting
$\Omega_2=0$ they become:
\begin{align}
\label{eq99}
&w_1^2 (k+2)^2(k+3)a_{k+3}+\Omega_1(k+1) a_{k+1}=0\;\;\; \mbox{for}\;\;\;k\geq1, \\
&12w_1^2a_3+\Omega_1a_1=0.
\end{align}
As before, there are three initial parameters that can freely be chosen, $a_0$, $a_1$, $a_2$.
Let us make the same choice as in Eq. (\ref{eq40}):
\begin{align}
\label{eq101}
\notag \tilde{L}_1(r) \;\;\;\mbox{with}\;\;\; a_0=1, a_1=0, a_2=0, \\
\tilde{L}_2(r) \;\;\;\mbox{with}\;\;\; a_0=0, a_1=1, a_2=0, \\
\notag \tilde{L}_3(r) \;\;\;\mbox{with}\;\;\; a_0=0, a_1=0, a_2=1.
\end{align}
The solutions of Eq. (\ref{eq98}) are denoted by tilde to distinguish them from the solutions in the
general case. The resulting functions can be expressed in terms of the generalized hypergeometric function
$\left(_pF_q\left[a_1,\ldots,a_p;b_1,\ldots,b_q;z\right]\right)$ and some elementary functions:
\begin{align}
\label{eq102}
&\tilde{L}_1(r)=1,\\
&\tilde{L}_2(r)=r\; _1F_2\left[\frac{1}{2};1,\frac{3}{2};-\frac{\Omega_1}{w_1^2}r^2\right], \\
&\tilde{L}_3(r)=r^2\; _2F_3\left[1,1;\frac{3}{2},\frac{3}{2},2;-\frac{\Omega_1}{w_1^2}r^2\right].
\end{align}
The recursion relation for the coefficients $b_k$ becomes:
\begin{align}
&2\Omega_1a_1+28w_1^2a_3+\Omega_1b_1+12w_1^2b_3=0,\\
\begin{split}
&2k\Omega_1a_k+2(k+1)(2k^2+4k+1)w_1^2a_{k+2}\\
&+k^2\Omega_1b_k+w_1^2k(k+1)^2(k+2)b_{k+2}=0.
\end{split}
\end{align}
At this point it is very useful to depart slightly from the previous approach and choose a little
less obvious initial conditions for the series $b_k$:
\begin{align}
b_0=0,\;\;\;b_1=\frac{4}{\pi}\left[\mbox{Log}\left(\frac{\sqrt{\Omega_1}}{2w_1}
\right)+\gamma-1\right ] , \;\;\;b_2=0.
\end{align}
With this choice the function $L_4(r)$ takes a very appealing form:
\begin{align}
\tilde{L}_4(r)&=\pi\, r\left[
Y_0\left(\frac{\sqrt{\Omega_1}}{w_1}r\right)H_{-1}\left(\frac{\sqrt{\Omega_1}}{w_1}r\right)\right.\\
&\left.+Y_1\left(\frac{\sqrt{\Omega_1}}{w_1}r\right)H_0\left(\frac{\sqrt{\Omega_1}}{w_1}r\right)
\right],
\end{align}
where $Y_\alpha$ is the Bessel function of the second kind and $H_\alpha$ is the Struve function, both of
the order $\alpha$. This completes the solution of Eq. (\ref{eq98}). In this particular case we
found a closed expression for $f_H(r)$ in terms of the known special functions, so that
the implementation and numerical realisation becomes significantly simpler. Since the initial
conditions for $f(r)$ in this special case are the same as in the general case, the solution of
(\ref{eq97}) takes the form:
\begin{align}
\label{eq76a}
\begin{split}
f(r)&=
\frac{\tilde{L}_1(r)}{w_1^2 \tilde{W}(1)}\int_0^r dr'\; r'\, U(r') \tilde{W}_1(r') \\
&-\frac{\tilde{L}_2(r)}{w_1^2 \tilde{W}(1)}\int_r^\infty dr'\; r'\, U(r') \tilde{W}_2(r') \\
&-\frac{\tilde{L}_3(r)}{w_1^2 \tilde{W}(1)}\int_r^\infty\ dr'\; r'\, U(r') \tilde{W}_3(r') \\
&+\frac{\tilde{L}_4(r)}{w_1^2 \tilde{W}(1)}\int_0^r dr'\; r'\, U(r') \tilde{W}_4(r'),
\end{split}
\end{align}
and the formulas for the derivatives are analogous to Eq. (\ref{eq77a})$-$(\ref{eq77c})

Whenever the $\Omega_2$ coefficient vanishes, the recursion relations established in the previous
sections are no longer correct. Equations for $\mbox{E}_{i}$, $i=1,5$, become a system of linear equations with
a singular coefficients matrix. To give an example how to circumvent this problem, let us derive an
analytical equation for $\frac{\partial f}{\partial w_3}$. In this special case $\mbox{E}_1$
and $\mbox{E}_2$ are:
\begin{align}
\label{eq102a}
\begin{split}
\mbox{E}_1=&-w_1^2\, r\, \frac{\partial f^{(4)}}{\partial w_3}-2 w_1^2 \,\frac{\partial
f^{(3)}}{\partial w_3}(r)
-\frac{\partial \Omega_1}{\partial w_3} r f''(r)\\
-&\Omega_1 r \frac{\partial f''}{\partial w_3}
-\frac{\partial \Omega_1}{\partial w_3} f'(r)
-\Omega_1 \frac{\partial f'}{\partial w_3}
-\frac{\partial \Omega_2}{\partial w_3} rf(r)\\
+&\frac{\partial U(r)}{\partial w_3}=0
\end{split}
\end{align}
\begin{align}
\label{eq102b}
\begin{split}
\mbox{E}_2&=w_1^2\, \frac{\partial f^{(4)}}{\partial w_3}+
\Omega_1 \frac{\partial f''}{\partial w_3}+
\frac{1}{2}\frac{\partial \Omega_1}{\partial w_3}f''(r)\\
&+\frac{1}{2}\frac{\partial \Omega_2}{\partial w_3}f(r)
+V(r)=0.
\end{split}
\end{align}
We form the combinations:
\begin{align}
\label{eq103}
\mbox{E}_3&=\frac{\partial}{\partial r}\left(\mbox{E}_1+r\, \mbox{E}_2\right), \\
\label{eq103aa}
\mbox{E}_4&=\frac{\partial}{\partial r}\left(r\, \mbox{E}_3-2\,\mbox{E}_1\right),
\end{align}
and solve Eqs. (\ref{eq102a})$-$(\ref{eq103aa}) for $\frac{\partial f'}{\partial w_3}$ instead of $\frac{\partial
f}{\partial w_3}$. The
result is:
\begin{align}
\label{eq103a}
\begin{split}
 \frac{\partial f'}{\partial w_3}=&\\
-& \frac{1}{2\Omega_1}
\left\lbrace
-2rV(r)-2\frac{\partial U(r)}{\partial w_3}
+\frac{\partial \Omega_2}{\partial w_3}rf(r)\right.\\
+&\left.\frac{\partial \Omega_1}{\partial w_3}[2f'(r)+r f''(r)]
\right\rbrace 
-\frac{w_1^2}{\Omega_1^2}
\left\lbrace
-8V'(r)-2rV''(r)\right.\\
-&\left.2\frac{\partial U''(r)}{\partial w_3}
+\frac{\partial \Omega_2}{\partial w_3}r f''(r)
+\frac{\partial \Omega_1}{\partial w_3}[2f^{(3)}(r)+rf^{(4)}(r)]
\right\rbrace
\end{split}
\end{align}
This result needs now to be formally integrated over $r$. The resulting integrals can be expanded as:
\begin{align}
\label{eq104}
\left[rf(r)\right]^{(-1)}=rf^{(-1)}(r)-f^{(-2)}(r),
\end{align}
where the superscript ${(-I)}$ was introduced to denote the $I$-fold integration over $r$ with the boundary condition
$f^{(-I)}(\infty)=0$ (the so-called antidifferentiation). The integrals of $f(r)$ over $r$ can be obtained
by the consecutive integration of Eq. (\ref{eq97}):
\begin{align}
\label{eq105}
f^{(-1)}(r)=\frac{1}{\Omega_1}\left[w_1^2r f''(r)+\Omega_1 r f(r)-U^{(-2)}(r)\right].
\end{align}
The result becomes:
\begin{widetext}
\begin{align}
\label{e106}
\notag \frac{\partial f}{\partial w_3}=&\\
-&\notag \frac{1}{2\Omega_1}
\left\lbrace
-2rV^{(-1)}(r)+2V^{(-2)}(r)-2\frac{\partial U^{(-1)}(r)}{\partial w_3}
+\frac{\partial \Omega_2}{\partial w_3}[rf^{(-1)}(r)-f^{(-2)}(r)]
+\frac{\partial \Omega_1}{\partial w_3}[f(r)+r f'(r)]
\right\rbrace \\
-& \frac{w_1^2}{\Omega_1^2}
\left\lbrace
-8V(r)-2rV'(r)+2V(r)-2\frac{\partial U'(r)}{\partial w_3}
+\frac{\partial \Omega_2}{\partial w_3}[r f'(r)-f(r)]
+\frac{\partial \Omega_1}{\partial w_3}[f''(r)+rf^{(3)}(r)]
\right\rbrace.
\end{align}
\end{widetext}
Antiderivatives of the functions $U(r)$ and $V(r)$ can all be obtained in an analytical way. For example, in the
special case $u_2=u_3=w_2=w_3=x$ mentioned earlier they are:
\begin{widetext}
\begin{align}
\label{eq107}
\notag\frac{\partial U^{(-1)}(r)}{\partial w_3}=&
\frac{w_1 e^{-r (w_1+2 x)}}{8 (w_1+2 x)} \left\lbrace-4 r w_1 e^{r
w_1} (w_1+2 x) \left[2 \mbox{Ei}(-r w_1)-e^{4 r x} \mbox{Ei}(-4 r x)-\mbox{Log} (4 r)\right.\right.\\
+&\left.\left.2 \mbox{Log} (w_1+2 x)-2 \mbox{Log} (w_1)-\mbox{Log} (x)\right]+4 e^{r w_1} \left[\gamma  r w_1(w_1+2 x)-4
x\right]+8 (w_1+2 x)\right\rbrace,
\end{align}
\begin{align}
\label{eq108}
\notag V^{(-1)}(r)=&
\frac{w_1 e^{-r (2 x+w_1)}}{2 r^2 (2 x+w_1)} \left\lbrace-r^2 w_1e^{r w_1} \left[e^{4 r x} (2 x+w_1) \mbox{Ei}(-4 r x)-2
(2 x+w_1) \mbox{Ei}(-r w_1)+w_1 \mbox{Log} (4 rx)\right.\right.\\ \notag
+&\left.\left.2 x \mbox{Log} (r x)+x \log (16)\right]+e^{r w_1} \left[-\gamma  r^2 w_1 (2 x+w_1)+4 r^2 w_1 (2 x+w_1)
\tanh ^{-1}\left(\frac{x}{x+w_1}\right)\right.\right. \\ +&\left.\left.4 rx+2\right]
-2 (r (2 x+w_1)+1)\right\rbrace,
\end{align}
\begin{align}
\label{eq109}
\notag \notag V^{(-2)}(r)=&\frac{w_1 e^{-r (4 x+w_1)}}{4 r x (2 x+w_1)} \left\lbrace e^{r (2 x+w_1)} \left[-r w_1^2e^{4
r x} \mbox{Ei}(-4 r x)+2 r w_1^2 e^{2 r x} \mbox{Ei}[-r (2 x+w_1)]\right.\right. \\ \notag
-&\left.\left. 2 rx w_1 e^{4 r x} \mbox{Ei}(-4 r x)+4 r x w_1 e^{2 r x} \mbox{Ei}[-r (2 x+w_1)]+r w_1^2
\mbox{Log} (4 r x)+\gamma  r w_1 (2 x+w_1)\right]\right. \\ \notag
+&\left.e^{r (2 x+w_1)}\left[2 r x w_1 \mbox{Log} (r x)+r x w_1\mbox{Log} (16)-4 r w_1 (2 x+w_1) \tanh
^{-1}\left(\frac{x}{x+w_1}\right)-4 x\right] \right.\\ 
-&\left.2 r w_1 (2 x+w_1) \mbox{Ei}(-r w_1) e^{r (2 x+w_1)}+4 x e^{2 rx}\right\rbrace.
\end{align}
\end{widetext}
In a similar way higher-order derivatives over the nonlinear parameters can be
calculated. One needs to use the expressions for $\mbox{E}_1$ and $\mbox{E}_2$ differentiated  the desired number of
times over
$u_2, u_3, w_2, w_3$ as a starting point and form the same combinations as in the above
example.

\subsection{Vanishing $\Omega_1$ coefficient}
\label{subsec:omega1}

Vanishing $\Omega_1$ is a by far less troublesome special case than the one considered in the previous subsection.
Conditions under which $\Omega_1$ vanishes are found by recasting it into a particular form
\begin{align}
\label{eq110}
\Omega_1=w_1^4-w_1^2\left(u_2^2+u_3^2+w_2^2+w_3^2\right)+(u_2^2-u_3^2)(w_3^2-w_2^2),
\end{align}
so we may solve $\Omega_1=0$ against $w_1^2$. The result is trivially found to be
$w_1^2=\frac{1}{2}\left( u_2^2+u_3^2+w_2^2+w_3^2 \pm \sqrt{\Delta} \right),$ 
with
$\Delta=\left(u_2^2+u_3^2+w_2^2+w_3^2\right)^2-4(u_2^2-u_3^2)(w_3^2-w_2^2)$
and we see that $\Omega_1$ vanishes after some coincidental choice of the nonlinear parameters defined by the above
equation rather than for some particular class of the integrals. In this special case the homogeneous differential
equation takes the form:
\begin{align}
\label{eq112}
w_1^2 r f_H^{(4)}(r)+2 w_1^2 f_H^{(3)}(r)+\Omega_2 rf_H(r)=0
\end{align}
This equation can be solved by using the recursion relations for the coefficients in the
series expansions derived in the general case, by setting $\Omega_1=0$ and recognizing the resulting series in terms of
the well-known special functions. Since we have already presented a detailed example of such a procedure, here we only
list the final equations in a convenient form. The solutions of Eq. (\ref{eq112}) are denoted
by double-tilde to distinguish them from the previous ones:
\begin{align}
\label{eq113}
&\overset{\approx}{L}_1(r)=\,_0F_3\left[\frac{1}{2},\frac{3}{4},\frac{3}{4};-\frac{\Omega_2}{256 w_1^2}r^4\right], \\
&\overset{\approx}{L}_2(r)=r\, _0F_3\left[\frac{3}{4},1,\frac{5}{4};-\frac{\Omega_2}{256 w_1^2}r^4\right], \\
&\overset{\approx}{L}_3(r)=r^2\, _0F_3\left[\frac{5}{4},\frac{5}{4},\frac{3}{2};-\frac{\Omega_2}{256 w_1^2}r^4\right],
\\
&\overset{\approx}{L}_4(r)=G^{20}_{04}\left(-\frac{\Omega_2}{256 w_1^2}r^4\right|\left. 
\begin{array}{c}
0 \\
\frac{1}{4},\frac{1}{4},0,\frac{1}{2}
\end{array}
 \right).
\end{align}
where 
$G^{mn}_{pq}\left(z |
\begin{array}{c}
a_1,\ldots,a_p \\
b_1,\ldots,b_q
\end{array}
\right)$
is the Meijer $G$-function. The solution of the inhomogeneous equation can now formally be written as:
\begin{align}
\label{eq115}
\begin{split}
f(r)&=
\frac{\overset{\approx}{L}_1(r)}{w_1^2 \overset{\approx}{W}(1)}\int_0^r dr'\; r'\, U(r') \overset{\approx}{W}_1(r') \\
&-\frac{\overset{\approx}{L}_2(r)}{w_1^2 \overset{\approx}{W}(1)}\int_r^\infty dr'\; r'\, U(r')
\overset{\approx}{W}_2(r') \\
&-\frac{\overset{\approx}{L}_3(r)}{w_1^2 \overset{\approx}{W}(1)}\int_r^\infty\ dr'\; r'\, U(r')
\overset{\approx}{W}_3(r') \\
&+\frac{\overset{\approx}{L}_4(r)}{w_1^2 \overset{\approx}{W}(1)}\int_0^r dr'\; r'\, U(r') \overset{\approx}{W}_4(r'), 
\end{split}
\end{align}
with the definitions of $\overset{\approx}{W}(r)$ and $\overset{\approx}{W}_i(r)$ analogous to Eq. (\ref{eq64}) and
(\ref{eq65}), respectively.

Since the recursion relations derived in the general case remain valid for $\Omega_1=0$ we can rewrite them as
they take much simpler form here, for instance:
\begin{align}
\label{eq116}
\begin{split}
\frac{\partial f}{\partial w_3}&=
\frac{1}{4\Omega_2} \left\lbrace
6V(r)+2rV'(r)+2\frac{\partial U'(r)}{\partial w_3}\right.\\
&+\frac{\partial \Omega_2}{\partial w_3}[f(r)+rf'(r)]
\left.-\frac{\partial \Omega_1}{\partial w_3}[f''(r)+rf^{(3)}(r)]
\right\rbrace, 
\end{split}
\end{align}
so that higher-order derivatives over the nonlinear parameters are calculated from
the recursion relation for the general case by putting $\Omega_1=0$ at the end of each recursive step.

\subsection{Vanishing $\Omega_1$ and $\Omega_2$ coefficients}
\label{sec:omega12}

Situation when $\Omega_1=0$ and $\Omega_2=0$ is quite rare since the conditions given in the two
previous subsections that make  $\Omega_1$ and $\Omega_2$ vanish must mutually be satisfied. This
occurs, for example, when $u_2=u_3=w_2=w_3=x$ and additionally $w_1=2x$. The homogeneous differential
equation has disarmingly simple four linearly independent solutions:
\begin{align}
\label{eq118}
&\check{L}_1(r)=1,\\
&\check{L}_2(r)=r,\\
&\check{L}_3(r)=r^2,\\
&\check{L}_4(r)=r\, \mbox{Log}(r)-r.
\end{align}
The above solutions were denoted by check mark to separate them from the previous ones. The Wronskian $W(r)$ and
the $W_i(r)$ determinants can be brought into  the following closed forms:
\begin{align}
&\check{W}(r)=-\frac{2}{r^2},\\
&\check{W}_1(r)=r,\\
&\check{W}_2(r)=2-2\, \mbox{Log}(r),\\
&\check{W}_3(r)=-\frac{1}{r},\\
&\check{W}_4(r)=-2,
\end{align}
so that the solution of the inhomogeneous differential equation takes the form:
\begin{align}
\begin{split}
f(r)&=\frac{1}{w_1^2}\left\lbrace
-\frac{1}{2}\int_0^r dr'\;r'^2U(r')
+r\int_r^\infty dr'\; r'U(r')\left[1-\mbox{Log}(r)\right] \right. \\ &
\left.-\frac{1}{2}r^2\int_r^\infty dr'\; U(r')
+r\left[\mbox{Log}(r)-1\right]\int_0^r dr'\; r' U(r')
\right\rbrace,
\end{split}
\end{align}
where for example in the case $u_2=u_3=w_2=w_3=x$, $w_1=2x$:
\begin{align}
\begin{split}
U(r)=&\frac{4 x e^{-4 r x}}{r^2} \left\lbrace 2 r^2 x^2 e^{2 r x} \left[e^{4 r x} \mbox{Ei}(-4 r
x)-2 \mbox{Ei}(-2 r x)\right.\right.\\
+&\left.\left.\mbox{Log} (r x)\right]+6 r x
+ e^{2 r x} [2 r x(\gamma  r x-1)-1]+1\right\rbrace.
\end{split}
\end{align}
However, even in such a simple case not all of the above integrals can be calculated fully
analytically, so we still need to struggle with the numerical integration. A little bit more
difficult is the differentiation of the master integral with respect to the nonlinear parameters. For
example, the derivative over $w_3$ is obtained from the special forms of the two identities which were
derived in the previous subsections:
\begin{align}
\begin{split}
\check{\mbox{E}}_1=&-w_1^2\, r\, \frac{\partial f^{(4)}}{\partial w_3}-2 w_1^2
\,\frac{\partial
f^{(3)}}{\partial w_3}(r)\\
-&\frac{\partial \Omega_1}{\partial w_3} r f''(r)
-\frac{\partial \Omega_1}{\partial w_3} f'(r)\\
-&\frac{\partial \Omega_2}{\partial w_3} rf(r)
+\frac{\partial U(r)}{\partial w_3}=0,
\end{split}
\end{align}
\begin{align}
\begin{split}
\check{\mbox{E}}_2=\,&w_1^2\, \frac{\partial f^{(4)}}{\partial w_3}+
\frac{1}{2}\frac{\partial \Omega_1}{\partial w_3}f''(r)\\
+&\frac{1}{2}\frac{\partial \Omega_2}{\partial w_3}f(r)+V(r)=0.
\end{split}
\end{align}
We take the combination $\left(\check{\mbox{E}}_1+r\,\check{\mbox{E}}_2\right)$ to cancel out the
term $\frac{\partial f^{(4)}}{\partial w_3}$ and solve the resulting equation against
$\frac{\partial f^{(3)}}{\partial w_3}$:
\begin{align}
\begin{split}
\frac{\partial f^{(3)}}{\partial w_3}=&
\frac{1}{2w_1^2}\left[
\frac{\partial U(r)}{\partial w_3}+rV(r)-\frac{\partial \Omega_1}{\partial w_3} f'(r)\right.\\
-&\left.\frac{1}{2}\frac{\partial \Omega_1}{\partial w_3} r f''(r)
-\frac{1}{2}\frac{\partial \Omega_2}{\partial w_3} rf(r)
\right].
\end{split}
\end{align}
This equation needs now to be antidifferentiated three times to give:
\begin{align}
\begin{split}
\frac{\partial f}{\partial w_3}&=
\frac{1}{2w_1^2}\left\lbrace
\frac{\partial U^{(-3)}(r)}{\partial w_3}+rV^{(-3)}(r)-3V^{(-4)}(r)\right. \\\
&-\frac{\partial\Omega_1}{\partial w_3} f^{(-2)}(r)
-\left.\frac{1}{2}\frac{\partial \Omega_1}{\partial w_3} [r f^{(-1)}(r)-3f^{(-2)}(r)]\right.\\
&-\left.\frac{1}{2}\frac{\partial \Omega_2}{\partial w_3} [r f^{(-3)}(r)-f^{(-4)}(r)]
\right\rbrace.
\end{split}
\end{align}
The antiderivatives of the master integral are obtained by consecutive antidifferentiation of the expression:
\begin{align}
\label{eq117}
\begin{split}
w_1^2 r f^{(4)}(r)+2 w_1^2 f^{(3)}(r)=U(r), \\ e.g. \; \; \;
f^{(-1)}(r)=rf(r)-\frac{U^{(-4)}(r)}{w_1^2}.
\end{split}
\end{align}
Higher-order derivatives with respect to the nonlinear parameters are obtained using the same
procedure albeit $\mbox{E}_1$ and $\mbox{E}_2$ need to be differentiated an arbitrary number of
times with respect to $u_2, u_3, w_2, w_3$ before putting $\Omega_1=0$ and $\Omega_2=0$.

\subsection{Vanishing $\Omega_1^2-4w_1^2\Omega_2$ coefficient}
\label{subsec:delta}

In the special case $\Omega_2=\left( \frac{\Omega_1}{2w_1} \right)^2$ two pairs of roots of the $\sigma$ polynomial,
Eqs. (\ref{eq19}) and (\ref{eq23}), lying on the same side of the complex plane coincide, so that $\sigma=w_1^2\left(
t^2+\frac{\Omega_1}{2w_1^2}\right)^2$. As a result, the homogeneous differential equation, Eq. (\ref{eq34}), can be
brought into the form:
\begin{align}
\label{eq118}
w_1^2\hat{A}\left[r\hat{A} f_H(r)\right]=0,
\end{align}
where $\hat{A}$ is a differential operator defined as:
\begin{align}
\label{eq119}
\hat{A}=\frac{\partial^2}{\partial r^2}-q^2,
\end{align}
with $q^2=-\frac{\Omega_1}{2w_1^2}$. Eq. (\ref{eq118}) can be solved by decomposing it into a system of two second-order
differential equations:
\begin{align}
\label{eq120}
w_1^2\hat{A}\left[rh(r)\right]=0,\\
\label{eq121}
\hat{A} f_H(r)=h(r).
\end{align}
The first of these equations has the form $rh''(r)+2h'(r)-rq^2h(r)=0$, so that the general solution is:
\begin{align}
h(r)=C_1 \frac{e^{qr}}{r}+C_2 \frac{e^{-qr}}{r},
\end{align}
and Eq. (\ref{eq121}) takes the form:
\begin{align}
\label{eq122}
f_H''(r)-q^2f_H(r)=C_1 \frac{e^{qr}}{r}+C_2 \frac{e^{-qr}}{r}.
\end{align}
The latter equation is solved with elementary methods. Finally we conclude that Eq. (\ref{eq118}) has four linearly
independent solutions that can be chosen as:
\begin{align}
\label{eq123}
\begin{split}
&e^{-qr},\;\;
e^{qr},\;\;
e^{-qr}\,\mbox{Ei}[2qr]-e^{qr}\,\mbox{Log}[2qr],\;\;\\
&\mbox{and}\;\; e^{qr}\,\mbox{Ei}[-2qr]-e^{-qr}\,\mbox{Log}[2qr].
\end{split}
\end{align}
The solution of the inhomogeneous differential equation takes the form analogous to Eq. (\ref{eq76}). Similarly, Eqs.
(\ref{eq77a})$-$(\ref{eq77c}) are the derivatives of the master integral with respect to $r$. 

Since $D_0=\Omega_1^2-4w_1^2\Omega_2$ appears in the denominator in nearly all recursion relations derived for the
general case they become invalid here. However, this problem can be circumvented by using the same trick as
in the $\Omega_2=0$ case, namely solving the system of Eqs. (\ref{eq102a})$-$(\ref{eq103aa}) with respect to
$\frac{\partial
f'}{\partial w_3}$ and performing consecutive antidifferentiations. Since the derivation is exactly the same as in the
subsection \ref{subsec:omega2} there is little point in repeating it here.

\section{Numerical examples}
\label{sec:numerical}

In this section we present results of calculations on the representative set of master
integrals with some hand-picked values of the nonlinear parameters. 
We implemented a general code that is able to
calculate the values of the master integral with arbitrarily chosen nonlinear parameters. The  code is
written in the C programming language and all the calculations were performed in the quadruple arithmetic precision
using the GCC Libquadmath library. Handful of the results presented here were
additionally
checked by using an independent program written in \emph{Mathematica} with the octuple arithmetic precision.
Comparison with the results obtained in the extended precision shows that calculations performed in
quadruple-precision, using $101$ points of the Tanh-Sinh quadrature~\cite{tanhsinh1,tanhsinh2} for
all numerical integrations,
gave an accuracy of at least long double precision (around 20 significant digits) and much better on
the average.

\begin{table}[t]
\label{table1}
\caption{Comparison of the values of the master integral with $u_2=w_2=1$, $u_3=w_3=2$, $w_1=1.0$
calculated according to Eqs. (\ref{eq118}) and (\ref{eq76}). Calculations performed for $r=2$. [k] denotes
10$^k$.}
\begin{ruledtabular}
\begin{tabular}{cD{.}{.}{1.24}}
$\mbox{expansion length}$ & \multicolumn{1}{c}{\mbox{$f(r)$ value}} \\
\hline
10 & 2.38 528 323 813 123 779 081 [-04]\\
20 & 2.39 121 157 642 870 208 323 [-04]\\
30 & 2.39 121 170 158 061 051 314 [-04]\\
40 & 2.39 121 170 158 262 980 140 [-04]\\
50 & 2.39 121 170 158 262 983 284 [-04]\\
60 & 2.39 121 170 158 262 983 284 [-04]\\
70 & 2.39 121 170 158 262 983 284 [-04]\\
75 & 2.39 121 170 158 262 983 284 [-04]\\
\hline
Eq. (\ref{eq76}) & 2.39 121 170 158 262 983 284 [-04] \\
\end{tabular}
\end{ruledtabular}
\end{table}

\begin{table}[t]
\label{table2}
\caption{Comparison of the values of the master integral with $u_2=w_2=1$, $u_3=w_3=2$, $w_1=1.5$
calculated according to Eqs. (\ref{eq118}) and (\ref{eq76}). Calculations performed for $r=2$. [k] denotes
10$^k$.}
\begin{ruledtabular}
\begin{tabular}{cD{.}{.}{1.24}}
$\mbox{expansion length}$ & \multicolumn{1}{c}{\mbox{$f(r)$ value}} \\
\hline
10 & 1.32 952 081 604 592 501 191 [-04]\\
20 & 1.63 128 202 122 143 112 608 [-04]\\
30 & 1.63 165 160 669 687 214 817 [-04]\\
40 & 1.63 165 195 079 171 515 995 [-04]\\
50 & 1.63 165 195 110 063 806 339 [-04]\\
60 & 1.63 165 195 110 091 457 699 [-04]\\
70 & 1.63 165 195 110 091 482 555 [-04]\\
75 & 1.63 165 195 110 091 482 578 [-04]\\
\hline
Eq. (\ref{eq76}) & 1.63 165 195 110 091 482 597 [-04] \\
\end{tabular}
\end{ruledtabular}
\end{table}

\begin{table}[t]
\centering
\label{table3}
\caption{Comparison of the values of the master integral with $u_2=w_2=1$, $u_3=w_3=2$, $w_1=2.0$
calculated according to Eqs. (\ref{eq118}) and (\ref{eq76}). Calculations performed for $r=2$. [k] denotes
10$^k$.}
\begin{ruledtabular}
\begin{tabular}{cD{.}{.}{1.24}}
$\mbox{expansion length}$ & \multicolumn{1}{c}{\mbox{$f(r)$ value}} \\
\hline
10 &-3.63 011 904 625 504 245 573 [-04]\\
20 & 1.07 042 192 138 901 654 774 [-04]\\
30 & 1.17 367 132 416 023 521 971 [-04]\\
40 & 1.17 538 021 409 679 753 524 [-04]\\
50 & 1.17 540 746 158 661 903 854 [-04]\\
60 & 1.17 540 789 464 360 697 759 [-04]\\
70 & 1.17 540 790 155 566 410 929 [-04]\\
75 & 1.17 540 790 165 897 003 951 [-04]\\
\hline
Eq. (\ref{eq76}) & 1.17 540 790 166 845 070 630 [-04] \\
\end{tabular}
\end{ruledtabular}
\end{table}

\begin{table}
\label{table4}
\caption{Comparison of the values of the master integral with $u_2=w_2=1$, $u_3=w_3=2$, $w_1=2.5$
calculated according to Eqs. (\ref{eq118}) and (\ref{eq76}). Calculations performed for $r=2$. [k] denotes
10$^k$.}
\begin{ruledtabular}
\begin{tabular}{cD{.}{.}{1.24}}
$\mbox{expansion length}$ & \multicolumn{1}{c}{\mbox{$f(r)$ value}} \\
\hline
10 &-3.96 382 454 835 359 866 193 [-03]\\
20 &-7.39 453 380 391 436 518 395 [-04]\\
30 &-3.92 253 435 711 155 687 731 [-05]\\
40 & 6.94 093 019 040 750 416 082 [-05]\\
50 & 8.55 501 340 987 912 083 454 [-05]\\
60 & 8.79 376 564 476 664 924 936 [-05]\\
70 & 8.82 922 222 853 990 351 131 [-05]\\
75 & 8.83 787 155 222 436 438 480 [-05]\\
\hline
Eq. (\ref{eq76}) & 8.83 545 586 039 834 446 778 [-05] \\
\end{tabular}
\end{ruledtabular}
\end{table}

\begin{table}
\centering
\label{table5}
\caption{Comparison of the values of the master integral with $u_2=w_2=1$, $u_3=w_3=2$, $w_1=3.0$
calculated according to Eqs. (\ref{eq118}) and (\ref{eq76}). Calculations performed for $r=2$. [k] denotes
10$^k$.}
\begin{ruledtabular}
\begin{tabular}{cD{.}{.}{1.24}}
$\mbox{expansion length}$ & \multicolumn{1}{c}{\mbox{$f(r)$ value}} \\
\hline
10 &-2.28 509 213 221 681 747 900 [-02]\\
20 &-2.90 220 438 815 452 281 770 [-02]\\
30 &-2.77 016 299 506 271 493 255 [-02]\\
40 &-2.54 662 563 339 713 840 590 [-02]\\
50 &-2.33 351 029 802 116 868 289 [-02]\\
60 &-2.14 729 270 210 488 323 595 [-02]\\
70 &-2.19 874 109 964 523 467 927 [-02]\\
75 &-2.19 299 042 315 798 720 906 [-02]\\
\hline
Eq. (\ref{eq76}) & 3.89 692 438 286 734 239 005 [-06] \\
\end{tabular}
\end{ruledtabular}
\end{table}

It is easy to verify that for the calculation of the master integral one can also use a
different procedure, based on the series expansion of $\exp(-w_1 r_{12})$ around $w_1=0$ under
the sign of the integral in Eq. (\ref{eq1}). Since the latter expansion is uniformly convergent for any
positive value of $w_1$ one can perform term by term integration what leads to the identity:
\begin{align}
\label{eq118a}
f_0(r;w_1)=
\sum_{n=0}^\infty f_n(r;0) \frac{(-1)^nw_1^n}{n!}.
\end{align}
The above series is convergent for any value of $w_1$ and gives exactly the same numerical result
as Eq. (\ref{eq76}). The prescription how to calculate the integrals with an arbitrary power of $r_{12}$
but $w_1=0$, $f_n(r;0)$, was recently presented by means of the open-ended recursion relation, \emph{c.f.} Eq. (48) of
Ref.~\cite{pachucki09}. Therefore, Eq. (\ref{eq118}) is an interesting
alternative to
the analytical equation derived in the previous subsection. It is worth considering in details how
fast the above series expansion converges for a given value of $w_1$ and how many terms are
necessary to obtain long double precision which one can easily get by using Eq. (\ref{eq76})
throughout. To make such a comparison possible, we implemented the mentioned recursion relation to
advance the power of $r_{12}$ in $f_n(r;0)$ as much as necessary in the \emph{Mathematica} package.
However, the first problem encountered was the numerical stability of this recursion. Although
the starting values for the recursion were computed in the octuple arithmetic precision, after $n=75$ steps
only few digits were estimated to be correct. Therefore, it is rather pointless to go beyond
this value of $n$. On the other hand, further extension of the arithmetic precision which is
already two times bigger than for calculations based on Eq. (\ref{eq76}) will slow down
the code dramatically and make it inferior to the numerical integration approach. In Tables 1-5
we present a comparison of the values of the master integral obtained according to the Eq. (\ref{eq118})
with different expansion lengths and obtained by numerical integration in Eq. (\ref{eq76}). We have
chosen representative values of the nonlinear parameters ($u_2=w_2=1$, $u_3=w_3=2$) which
were kept fixed and we have varied the value of $w_1$ to examine the behaviour of the expansion Eq.
(\ref{eq118}) with increasing $w_1$. We see that for $w_1=1$ series expansion defined by Eq.
(\ref{eq118}) converges fast and smoothly towards the correct value and only a few tens of terms are
necessary to obtain the long double precision result. An even better behaviour is met for lower values of $w_1$.
However, when $w_1$ increases beyond one the convergence of Eq. (\ref{eq118}) deteriorates and
even for $w_1=1.5$ as much as 75 terms of the expansion are not enough to obtain a reliable precision
of 21 significant digits. For $w_1=2.5$ only three significant digits are recovered after 75 terms
and for $w_1=3.0$ the series converges so badly that no useful information about the value of the master
integral is obtained after 75 terms. Moreover, the result for $w_1=3.0$ is clearly wrong since
by a simple inspection of Eq. (\ref{eq1}) we observe that the value of the master integral is always
positive. 

Although a simple comparison provided in the above clearly shows that the numerical integration approach is superior
compared to the series expansion method, we must admit that the numerical integration has its own
problems. They appear for small values of $w_1$, say, lower than $0.2$. In this regime, the integrands in Eq.
(\ref{eq76})
vanish slowly and significant contribution to the value of the master integral comes from the large $r'$ in integration
over the interval $[r,+\infty[$. The treatment of such situations requires an efficient matching of the series expansion
around $0$ with the asymptotic expansions of the functions $L_i(r)$. Moreover, for large $r'$ accurate calculation of
$W_i(r')$ becomes difficult because significant loss of digits occurs due to the subtraction of two near-equal numbers.
Therefore, we believe that the most efficient method of calculation of the master integral will be a suitable union of
two algorithms described here. Series expansion is to be used for small $w_1$ where it converges fast and only a
handful of terms is required to obtain desired accuracy. For larger $w_1$ numerical integration is superior and
is able to provide arbitrary accuracy. In Table 6 we additionally listed values of the master integral with some
combinations of the nonlinear parameters corresponding to the general case without comparing them to the series
expansion method.

Special cases of the master integral were implemented separately taking advantage of the fact that $L_i(r)$ functions
are expressed in terms of known special functions. All necessary special functions were implemented using Chebyshev
interpolation method. In Table 7 we give examples of the values of the master integral in one important special case
corresponding to the exponential version of the symmetric James-Coolidge basis set, namely $u_2=u_3=w_2=w_3=x$ with an
arbitrary value of $w_1$ and $x$ (vanishing $\Omega_2$ coefficient).

\begin{table}
\label{table6}
\caption{Examples of the values of the master integral with $u_2=w_2=x$, $u_3=w_3=y$ calculated according to Eq.
(\ref{eq76}). The symbol [k] denotes the powers of 10, 10$^k$.}
\begin{ruledtabular}
\begin{tabular}{ccccD{.}{.}{1.24}}
$w_1$ & $x$ & $y$ & $r$ & \multicolumn{1}{c}{\mbox{$f(r)$ value}} \\
\hline
2.0 & 2.0 & 3.0 & 2.0 & 1.20 162 929 654 132 118 557 [-06]\\
2.0 & 2.0 & 3.0 & 6.0 & 1.85 690 183 473 424 018 070 [-15]\\
3.0 & 2.0 & 3.0 & 4.0 & 2.54 651 823 562 870 024 818 [-11]\\
3.0 & 2.0 & 3.0 & 1.0 & 1.27 020 704 100 393 024 656 [-04]\\
2.0 & 2.5 & 1.5 & 5.0 & 5.46 845 126 485 930 791 142 [-11]\\
2.0 & 2.5 & 1.5 & 10.0 & 1.23 085 262 413 134 222 873 [-11]\\
5.0 & 1.0 & 3.5 & 1.0 & 1.34 862 879 254 745 038 343 [-04]\\
5.0 & 1.0 & 3.5 & 3.0 & 1.12 445 814 382 106 648 815 [-08]\\
8.0 & 2.0 & 4.0 & 1.0 & 1.12 012 736 029 631 067 277 [-05]\\
8.0 & 2.0 & 4.0 & 0.5 & 2.73 482 762 158 894 864 378 [-04]\\
\end{tabular}
\end{ruledtabular}
\end{table}

\begin{table}
\label{table7}
\caption{Examples of the values of the master integral with $u_2=w_2=u_3=w_3=x$, calculated according to Eq.
(\ref{eq76a}). The symbol [k] denotes the powers of 10, 10$^k$.}
\begin{ruledtabular}
\begin{tabular}{cccD{.}{.}{1.24}}
$w_1$ & $x$ & $r$ & \multicolumn{1}{c}{\mbox{$f(r)$ value}} \\
\hline
1.0 & 1.0 & 1.0 & 1.70 312 528 092 841 122 796 [-02]\\
1.0 & 2.0 & 1.0 & 1.01 402 211 417 374 426 756 [-03]\\
1.0 & 3.0 & 1.0 & 8.00 415 011 434 536 226 469 [-05]\\
3.0 & 1.0 & 1.0 & 5.95 180 137 043 458 120 693 [-03]\\
3.0 & 2.0 & 1.0 & 4.27 403 013 549 289 339 748 [-04]\\
3.0 & 3.0 & 1.0 & 3.72 976 454 960 392 088 836 [-05]\\
1.0 & 1.0 & 6.0 & 7.05 646 817 317 973 364 415 [-07]\\
1.0 & 2.0 & 6.0 & 1.84 467 129 942 180 429 118 [-12] \\
1.0 & 3.0 & 6.0 & 6.55 011 946 233 785 000 932 [-18]\\
3.0 & 1.0 & 6.0 & 1.43 442 768 276 386 167 206 [-07]\\
3.0 & 2.0 & 6.0 & 4.47 644 083 118 078 530 669 [-13]\\
3.0 & 3.0 & 6.0 & 1.76 871 016 217 927 283 912 [-18]\\
\end{tabular}
\end{ruledtabular}
\end{table}

\section{Outline for the future}
\label{sec:conclusions}
In this paper we introduced a new explicitly correlated basis set for 
state-of-the-art \emph{ab-initio} calculations on diatomic molecules and reported analytical 
formulas ready to apply for all molecular integrals appearing in the 
non-relativistic calculations. While a  physical application of the theory 
presented here will be reported soon, we would like to stress that our 
theoretical results will find several important applications.

First of all, the Slater geminal basis is expected to improve the convergence
of molecular calculations on two-electron diatomic molecules by several orders
of magnitude. With the advent of new experimental tools that allow measurements
of the dissociation energy of H$_2$ with an astonishing accuracy of 10$^{-4}$
cm$^{-1}$ \cite{merkt09} and with the announcements that this level of accuracy
will be improved by two orders of magnitude, new molecular calculations will
be necessary to reproduce the experimental data. Especially important in this
respect will be the calculation of the relativistic integrals in the basis
of the Slater geminals. We expect that the accuracy of the relativistic 
corrections reported in Ref. \cite{piszczatowski09}, computed in the basis
of explicitly correlated Gaussian geminals, can be improved by a few orders
of magnitude. Also the QED effects could be accounted for in a more accurate
way, to produce not only state-of-the-art estimates of the dissociation
energy, but also of the rotational and vibrational spacings \cite{komasa11}.

The second important application of the theory presented above is the
numerical calculation of the integrals in the basis set of Slater orbitals
for diatomic molecules. While the theoretical background was
introduced by Pachucki in 2009 \cite{pachucki09}, his algorithms for certain
classes of integrals turned out to be inefficient for practical implementation. Using the geminal
recursion relations and putting the exponent in the correlation factor
$w_1$ equal to zero, one obtains much simpler and numerically more convenient
recursion relations.

Also worth mentioning are the calculations of the relativistic integrals
in the basis set of the Slater orbitals for diatomic molecules. At present,
no {\em ab initio} program for molecular calculations has available all 
integrals appearing in the Breit-Pauli theory, even in the Gaussian basis
set. We plan to apply our theory to the expressions for the most difficult
class involving the $r_{12}^{-2}$ factor, and perform actual calculations with
just one numerical integration in one dimension. In this way, accurate 
calculations of the fine and hyperfine structure of diatomics will become possible.

Finally, the basis set of the Slater geminals can be used in the explicitly 
correlated MBPT/CC theories, thus greatly improving the accuracy of the
present approaches based on the Gaussian orbitals and linear or exponential
correlation factors.

\begin{acknowledgments}
We would like to thank Professors Bogumi\l\ Jeziorski and Krzysztof Pachucki 
for many useful discussions, and for reading and commenting on the manuscript.
ML acknowledges the Polish Ministry of Science and Higher Education for the support
through the project ``Diamentowy Grant'', number DI2011 012041. RM was supported by the Polish Ministry of Science
and Higher Education, grant NN204 182840.

\end{acknowledgments}

\appendix

\section{Matrix elements of the kinetic energy operator}
\label{app:appa}

Let us consider the two-center two-electron Sch\"{o}dinger Hamiltonian:
\begin{align}
\hat{H}=-\frac{1}{2}\nabla_1^2-\frac{1}{2}\nabla_2^2-\frac{Z_A}{r_{1A}}-\frac{Z_A}{r_{2A}}
-\frac{Z_B}{r_{1B}}-\frac{Z_B}{r_{2B}}+\frac{1}{r_{12}}+\frac{1}{r},
\end{align}
where $Z_K$ denotes the nuclear charge of the nucleus $K$ and the notation for the other quantities is the same as in Eq. (\ref{eq0a}).
The basis functions are of the form (\ref{eq0}). The overlap integrals between these basis functions and the matrix
elements of the nuclear attraction and electronic repulsion operators are obviously expressed through the integrals from
the family (\ref{eq0a}) with proper powers of $r_{iK}$ and $r_{12}$. The only difficulty is to express the matrix
elements of the kinetic energy operator through the integrals (\ref{eq0a}). Let us introduce a
shorthand notation that will be used throughout this Appendix:
\begin{align}
\begin{split}
|ijkln\rangle&=
r_{1A}^ir_{1B}^jr_{2A}^kr_{2B}^lr_{12}^n
e^{-u_3 r_{1A}-u_2 r_{1B}-w_2 r_{2A}-w_3 r_{2B}-w_1 r_{12}}\\
&=\varphi_{ij}(1)\varphi_{kl}(2)r_{12}^n e^{-w_1 r_{12}}.
\end{split}
\end{align}
Our derivation was inspired by the procedure given by
Ko\l os
\emph{et al.}~\cite{kolos60a,kolos60b}. In fact their result is the $w_1=w_1'=0$ limit
of our equation. The matrix element is transformed as:
\begin{align}
\label{a4}
\begin{split}
&\langle ijkln| -\nabla_1^2 |i'j'k'l'n'\rangle=
\int d^3r_2 \;\varphi_{kl}(2)\,\varphi_{k'l'}(2)\\
&\times\int d^3r_1\; \nabla_1 \left[ \varphi_{ij}(1)r_{12}^n e^{-w_1 r_{12}}\right]\cdot 
\nabla_1 \left[ \varphi_{i'j'}(1)r_{12}^{n'} e^{-w_1' r_{12}}\right],
\end{split}
\end{align}
where the Green's theorem was used. Let us now consider only the integration over the coordinates of
the first electron. For simplicity we will consider the case $n=n'=0$. The general form of the
matrix element will then be obtained by differentiating $n$ times with respect to $w_1$ and $n'$
times with respect to $w_1'$, and multiplying by the factor $(-1)^{n+n'}$.
The latter integral is rewritten as:
\begin{align}
\label{a5}
\begin{split}
&\int d^3r_1\; \nabla_1 \left[ \varphi_{ij}(1) e^{-w_1 r_{12}}\right]\cdot 
\nabla_1 \left[ \varphi_{i'j'}(1) e^{-w_1' r_{12}}\right]\\
&=\int d^3r_1\; \left\lbrace
[\nabla_1\varphi_{ij}(1)][\nabla_1\varphi_{i'j'}(1)]e^{
-(w_1+w_1')r_{12}}\right.\\
&\left.+w_1w_1'\varphi_{ij}(1)\varphi_{i'j'}(1)e^{-(w_1+w_1')
r_{12}}
\right.\\
&\left.
+\frac{1}{w_1+w_1'}\left(
w_1'[\nabla_1\varphi_{ij}(1)]\varphi_{i'j'}(1) \right.\right.\\
&\left.\left.+w_1\varphi_{ij}(1)[\nabla_1\varphi_{i'j'}(1)]\right)
[\nabla_1 e^{-(w_1+w_1') r_{12}}]
\right\rbrace\\
&=\int d^3r_1\; \left\lbrace
w_1w_1'\varphi_{ij}(1)\varphi_{i'j'}(1)e^{-(w_1+w_1')
r_{12}}
\right.\\
&\left.
-\frac{1}{w_1+w_1'}\left(
w_1'[\Delta_1\varphi_{ij}(1)]\varphi_{i'j'}(1) \right.\right.\\
&\left.\left.+w_1\varphi_{ij}(1)[\Delta_1\varphi_{i'j'}(1)]\right)
e^{-(w_1+w_1') r_{12}}
\right\rbrace,
\end{split}
\end{align}
where the Green's theorem was used in the last step to remove the gradient operator working  on
terms containing the $r_{12}$ factor. The equation for the Laplacian of the $\varphi_{ij}(1)$ is
rather straightforward to derive and the final result is:
\begin{align}
\label{a6}
\begin{split}
\Delta_1\varphi_{ij}(1)&=
i(i+j+1)\varphi_{i-2,j}(1)+j(i+j+1)\varphi_{i,j-2}(1)\\
&+\left(u_3^2+u_2^2\right)\varphi_{ij}(1)
-u_3\left(i+\frac{j}{2}+2\right)\varphi_{i-1,j}(1)\\
&-u_2\left(j+\frac{i}{2}+2\right)\varphi_{i,j-1}(1)-r^2ij\varphi_{i-2,j-2}(1)\\
&+\frac{1}{2}iu_2r^2 \varphi_{i-2,j-1}(1)+\frac{1}{2}ju_3r^2\varphi_{i-1,j-2}(1)\\
&+u_2u_3\left[\varphi_{i+1,j-1}(1)+\varphi_{i-1,j+1}(1)-r^2\varphi_{i-1,j-1}(1)\right]\\
&-\frac{1}{2}iu_2\varphi_{i-2,j+1}(1)-\frac{1}{2}ju_3\varphi_{i+1,j-2}(1)
\end{split}
\end{align}
By inserting the formulas (\ref{a5}) and (\ref{a6}) into Eq. (\ref{a4}) one arrives at:
\begin{widetext}
\begin{align}
\label{a7}
\begin{split}
&\langle ijkl0| -\nabla_1^2 |i'j'k'l'0\rangle=
w_1w_1'\langle ijkl0|i'j'k'l'0\rangle\\
&-\frac{w_1'}{w_1+w_1'}\left[
i(i+j+1)\langle i-2,jkl0|i'j'k'l'0\rangle+
j(i+j+1)\langle ij-2,kl0|i'j'k'l'0\rangle\right.\\
&\left.+\left(u_3^2+u_2^2\right)\langle ijkl0|i'j'k'l'0\rangle-
u_3\left(i+\frac{j}{2}+2\right)\langle i-1,jkl0|i'j'k'l'0\rangle-
u_2\left(j+\frac{i}{2}+2\right)\langle ij-1,kl0|i'j'k'l'0\rangle\right.\\
&\left.-r^2ij\langle i-2,j-2,kl0|i'j'k'l'0\rangle
+\frac{1}{2}iu_2r^2 \langle i-2,j-1,kl0|i'j'k'l'0\rangle
+\frac{1}{2}ju_3r^2 \langle i-1,j-2,kl0|i'j'k'l'0\rangle\right.\\
&\left. +u_2u_3\langle i+1,j-1,kl0|i'j'k'l'0\rangle
+u_2u_3\langle i-1,j+1,kl0|i'j'k'l'0\rangle
-r^2u_2u_3\langle i-1,j-1,kl0|i'j'k'l'0\rangle\right.\\
&\left.-\frac{1}{2}iu_2 \langle i-2,j+1,kl0|i'j'k'l'0\rangle
-\frac{1}{2}ju_3 \langle i+1,j-2,kl0|i'j'k'l'0\rangle
\right]\\
&-\frac{w_1}{w_1+w_1'}\left[
i'(i'+j'+1)\langle ijkl0|i'-2,j'k'l'0\rangle+
j'(i'+j'+1)\langle ij,kl0|i'j'-2,k'l'0\rangle\right.\\
&\left.+\left(u_3'^2+u_2'^2\right)\langle ijkl0|i'j'k'l'0\rangle-
u_3'\left(i'+\frac{j'}{2}+2\right)\langle ijkl0|i'-1,j'k'l'0\rangle-
u_2'\left(j'+\frac{i'}{2}+2\right)\langle ijkl0|i'j'-1,k'l'0\rangle\right.\\
&\left.-r^2i'j'\langle ijkl0|i'-2,j'-2,k'l'0\rangle
+\frac{1}{2}i'u'_2r^2 \langle ijkl0|i'-2,j'-1,k'l'0\rangle
+\frac{1}{2}j'u'_3r^2 \langle ijkl0|i'-1,j'-2,k'l'0\rangle\right.\\
&\left. +u_2'u_3'\langle ijkl0|i'+1,j'-1,k'l'0\rangle
+u_2'u_3'\langle ijkl0|i'-1,j'+1,k'l'0\rangle
-r^2u_2'u_3'\langle ijkl0|i'-1,j'-1,k'l'0\rangle\right.\\
&\left.-\frac{1}{2}i'u_2' \langle ijkl0|i'-2,j'+1,k'l'0\rangle
-\frac{1}{2}j'u_3' \langle ijkl0|i'+1,j'-2,k'l'0\rangle
\right],
\end{split}
\end{align}
\end{widetext}
where the notation $\langle ijkl0|i'j'k'l'0\rangle$ was used to designate ordinary overlap
integrals which belong to the integral family (\ref{eq0a}).
The above equation is already symmetric with respect to the interchange of primed and non-primed indices.
It is noteworthy that the singularity appearing when $w_1=w_1'=0$ is only apparent. It
can easily be removed by taking first the limit $w_1' \rightarrow w_1$ and then putting
$w_1=0$. 
To obtain the most general matrix element of the kinetic energy operator, $\langle ijkln|
-\nabla_1^2 |i'j'k'l'n'\rangle$, one needs to differentiate Eq. (\ref{a7}) with respect to $w_1$
and $w_1'$. This
differentiation is easily carried out explicitly with any symbolic mathematical program. The resulting
formulas are quite compact and make their direct implementation straightforward. On the other hand,
by multiplying both sides of Eq. (\ref{a7}) by $w_1+w_1'$ and performing
consecutive differentiation, it is quite easy to derive a recursion relation that connects the values
of $\langle ijkln|-\nabla_1^2 |i'j'k'l'n'\rangle$ with different $n$ and $n'$. However, this recursion
is inherently unstable with the increasing $n$ and $n'$, and can be considered inferior to the
approach based on the analytical equations, at least for larger $n$ and $n'$.

\section{Analytical formulas for the functions $W(r)$ and $V(r)$}
\label{app:appb}
In this Appendix we list explicit analytical formulas for the functions $W(r;w_1,u_2,u_3,w_2,w_3)$
and $V(r;w_1,u_2,u_3,w_2,w_3)$ which are defined as the inverse Laplace transforms in Eqs. (\ref{eq33})
and (\ref{eq33a}), respectively. We would like to stress that in the formulas given below all the terms proportional
to the Dirac delta distribution or its derivatives were omitted. This can be done because for $r>0$ these terms
never contribute to the values of the master integral derivatives calculated from the recursion
relations presented in the text. However, the missing terms can be recovered by taking the
Laplace transform of the listed equations and comparing with the proper analogue of Eq. (\ref{eq19a}). This
can easily be done by using any symbolic mathematical package.

The function $W(r)$ takes the form:
\begin{widetext}
\begin{align}
W(r)=\sum_{i=1}^6 W_i(r)
\end{align}
with
\begin{align}
&W_1(r)=w_1 e^{-r (u_2+w_1+w_2)}
\left(\frac{1}{r^2}+\frac{u_2+w_1+w_2}{r}\right),\\
&W_2(r)=w_1 e^{-r (u_3+w_1+w_3)}
\left(\frac{1}{r^2}+\frac{u_3+w_1+w_3}{r}\right),
\end{align}
\begin{align}
\begin{split}
W_3(r)&=\frac{w_1 w_2(u_2^2-w_3^2)
 +u_2^2 w_2^2-u_3^2w_3^2}{(w_1+w_2)^2-w_3^2}
\frac{1}{r}\left[e^{-r(u_2+w_2+w_1)}-e^{-r(u_2+w_3)}\right]\\
&+\frac{w_1(w_1+w_2)}{(w_1+w_2)^2-w_3^2}\frac{1}{r^3}
\left\lbrace
-e^{-r(u_2+w_3)}\left[ 2+2r(u_2+w_3)+r^2(u_2+w_3)^2 \right]
\right.\\
&+\left.
e^{-r(u_2+w_2+w_1)}\left[ 2+2r(u_2+w_2+w_1)+r^2(u_2+w_2+w_1)^2 \right]
\right\rbrace,
\end{split}
\end{align}
\begin{align}
\begin{split}
W_4(r)&=\frac{w_1 w_3(u_3^2-w_2^2)
 +u_3^2 w_3^2-u_2^2w_2^2}{(w_1+w_3)^2-w_2^2}
\frac{1}{r}\left[e^{-r(u_3+w_3+w_1)}-e^{-r(u_3+w_2)}\right]\\
&+\frac{w_1(w_1+w_3)}{(w_1+w_3)^2-w_2^2}\frac{1}{r^3}
\left\lbrace
-e^{-r(u_3+w_2)}\left[ 2+2r(u_3+w_2)+r^2(u_3+w_2)^2 \right]
\right.\\
&+\left.
e^{-r(u_3+w_3+w_1)}\left[ 2+2r(u_3+w_3+w_1)+r^2(u_3+w_3+w_1)^2 \right] 
\right\rbrace,
\end{split}
\end{align}
\begin{align}
\begin{split}
W_5(r)&=\frac{w_1u_2(w_2^2-u_3^2)+u_2^2 w_2^2-u_3^2w_3^2}{(u_2+w_1)^2-u_3^2}
\frac{1}{r}\left[ e^{-r(u_2+w_2+w_1)}-e^{-r(u_3+w_2)} \right]\\
&+\frac{u_2^2-u_3^2+u_2w_1}{(u_2+w_1)^2-u_3^2}\frac{1}{r^3}
\left\lbrace
e^{-r(u_3+w_2)}\left[ 2+2r(u_3+w_2)+r^2(u_3+w_2)^2 \right] 
\right. \\
&\left.
-e^{-r(u_2+w_2+w_1)}\left[ 2+2r(u_2+w_2+w_1)+r^2(u_2+w_2+w_1)^2 \right] 
\right\rbrace,
\end{split}
\end{align}
\begin{align}
\begin{split}
W_6(r)&=\frac{w_1u_3(u_2^2-w_3^2)+u_2^2 w_2^2-u_3^2w_3^2}{-(u_2+w_1)^2+u_2^2}
\frac{1}{r}\left[ e^{-r(u_3+w_3+w_1)}-e^{-r(u_2+w_3)} \right]\\
&+\frac{u_2^2-u_3^2-u_3w_1}{-(u_3+w_1)^2+u_2^2}\frac{1}{r^3}
\left\lbrace
e^{-r(u_2+w_3)}\left[ 2+2r(u_2+w_3)+r^2(u_2+w_3)^2 \right] 
\right. \\
&\left.
-e^{-r(u_3+w_3+w_1)}\left[ 2+2r(u_3+w_3+w_1)+r^2(u_3+w_3+w_1)^2 \right] 
\right\rbrace,
\end{split}
\end{align}
\end{widetext}
Let us denote the permutation $u_2\leftrightarrow w_3$, $u_3\leftrightarrow w_2$ by $P_{12}$ and the permutation
$u_2\leftrightarrow u_3$, $w_2 \leftrightarrow w_3$ by $P_{AB}$. The reason for adopting such a notation becomes clear when
one considers the symmetries of the master integral. Calculation of $W(r)$ is simplified by the following relations:
\begin{align}
\begin{split}
P_{AB}W_1(r)=W_2(r), \\
P_{AB}W_3(r)=W_4(r), \\
P_{AB}W_5(r)=W_6(r),
\end{split}
\end{align}
so that the programming effort is halved. Further simplifications occur after observing,
for instance, that:
\begin{align}
\begin{split}
\frac{e^{-r(u_3+w_2)}}{r^3}\left[ 2+2r(u_3+w_2)+r^2(u_3+w_2)^2 \right]=\\
\frac{\partial^2}{\partial r^2}\left[ \frac{e^{-r(u_3+w_2)}}{r} \right]
\end{split}
\end{align}
so that in the implementation one can concentrate on the calculation of the derivatives of the quantities like $e^{-ar}/r$
since any derivative of $W(r)$ with respect to the nonlinear parameters and $r$ is expressed through them.

Explicit form of $V(r)$ is expressed conveniently as:
\begin{align}
V(r)=V_1(r)+V_2(r)+V_3(r)+\bar{c}_1V_4(r)+\bar{c}_2 V_5(r),
\end{align}
\begin{widetext}
with
\begin{align}
\begin{split}
V_1(r)&=\frac{e^{-r(u_3+w_2+w_1)}}{r}\left(w_1^2-u_3^2-u_3w_3\right)
+\frac{e^{-r(u_3+w_2)}}{r}\left(-w_1^2+u_3^2-u_2w_3\right)\\
&+u_3w_3\frac{e^{-r(u_2+w_2)}}{r}+u_2w_3\frac{e^{-r(u_2+w_2+w_1)}}{r},
\end{split}
\end{align}
\begin{align}
\begin{split}
V_2(r)&=\frac{w_1 w_3(u_3^2-w_2^2)
 +u_3^2 w_3^2-u_2^2w_2^2}{(w_1+w_3)^2-w_2^2}\frac{1}{r}
\left[e^{-r(u_3+w_3+w_1)}-e^{-r(u_3+w_2)}\right]\\
&+\frac{w_1(w_1+w_3)}{(w_1+w_3)^2-w_2^2}\frac{1}{r^3}
\left\lbrace
-e^{-r(u_3+w_2)}\left[ 2+2r(u_3+w_2)
+r^2(u_3+w_2)^2 \right]\right.\\
&+\left.e^{-r(u_3+w_3+w_1)}\left[ 2+2r(u_3+w_3+w_1)+ r^2(u_3+w_3+w_1)^2 \right] 
\right\rbrace,
\end{split}
\end{align}
\begin{align}
\begin{split}
V_3(r)&=\frac{w_1 w_3(w_2^2-u_3^2)
 +w_2w_3(u_3^2-u_2^2+w_1^2)}{(w_1+w_2)^2-w_3^2}
\frac{1}{r}\left[e^{-r(u_2+w_2+w_1)}-e^{-r(u_2+w_3)}\right]\\
&-\frac{w_1w_3}{(w_1+w_2)^2-w_3^2}\frac{1}{r^3}
\left\lbrace
-e^{-r(u_2+w_3)}\left[ 2+2r(u_2+w_3)+r^2(u_2+w_3)^2 \right]
\right.\\
&\left.
e^{-r(u_2+w_2+w_1)}\left[ 2+2r(u_2+w_2+w_1)+r^2(u_2+w_2+w_1)^2 \right]
\right\rbrace,
\end{split}
\end{align}
\begin{align}
\begin{split}
V_4(r)=&e^{-r \left(u_2+w_3\right)}
\left\lbrace-\text{Ei}\left[-r\left(w_1+u_3-u_2\right)\right]+\text{Ei}
\left[-r\left(u_3-u_2-w_3+w_2\right)\right]\right.\\
-&\left.\text{Ei}\left[-r \left(w_1-w_3+w_2\right)\right]\right\rbrace
-e^{r \left(u_2+w_3\right)} \text{Ei}\left[-r
\left(u_2+u_3+w_2+w_3\right)\right]\\ 
+&e^{-r
\left(u_2+w_3\right)}\mbox{Log}\left|\frac{\left(w_1+w_2-w_3\right)\left(w_1-u_2+u_3\right)\left(u_2+u_3+w_2+w_3\right)}
{
\left(w_1+w_2+w_3\right)
\left(u_2+u_3+w_1\right)
\left(u_2-u_3-w_2+w_3\right)}\right|,
\end{split}
\end{align}
\begin{align}
V_5(r)=& e^{r \left(w_3-u_2\right)}\text{Ei}\left[-r \left(w_1+w_2+w_3\right)\right]+e^{r
\left(u_2-w_3\right)} \text{Ei}\left[-r\left(w_1+u_2+u_3\right)\right],
\end{align}
and
\begin{align}
\bar{c}_1=\frac{1}{2}\left[ u_2\left(w_1^2-w_2^2+w_3^2\right)+w_3\left(u_2^2-u_3^2+w_1^2\right) \right],\\
\bar{c}_2=\frac{1}{2}\left[ u_2\left(w_1^2-w_2^2+w_3^2\right)-w_3\left(u_2^2-u_3^2+w_1^2\right) \right].
\end{align}
\end{widetext}
We see that the form of $V_2(r)$ and $V_3(r)$ is analogous to the $W_{i}(r)$, $i=3,6$. Similarly, $V_4(r)$ and $V_5(r)$
are
expressed through the same combinations of functions as $U_3(r)$ and $U_2(r)$, respectively. The only difference is
that some terms contribute with the sign reversed. Since the form of $U_1(r)$ is relatively simple and straightforward
to implement, arbitrary derivatives of the function $V(r)$ can be computed by using a proper union of the algorithms for
$U(r)$ and $W(r)$ functions.

\section{Auxillary recursion relations}
\label{app:appc}
In this Appendix we list formulas for the derivatives of $\frac{\partial f}{\partial w_3}$ and $\frac{\partial f}{\partial w_1}$ over $r$
up to the third-order which result from the solution of the set of equations
for $\mbox{E}_{i}$ and $\overline{\mbox{E}}_i$, $i=1,5$. 
Higher-order derivatives can be obtained by successive differentiation of the geminal differential equation:
\begin{widetext}
\begin{align}
\begin{split}
\frac{\partial f'}{\partial w_3}&=\\
&\frac{\Omega_1}{2\left(4w_1^2\, \Omega_2-\,\Omega_1^2\right)}
\left\lbrace
-2rV(r)-2\frac{\partial U(r)}{\partial w_3}
+\frac{\partial \Omega_2}{\partial w_3}rf(r)
+\frac{\partial \Omega_1}{\partial w_3}[2f'(r)+r f''(r)]
\right\rbrace \\
&+\frac{w_1^2}{4w_1^2\, \Omega_2-\Omega_1^2}
\left\lbrace
-8V'(r)-2rV''(r)-2\frac{\partial U''(r)}{\partial w_3}
+\frac{\partial \Omega_2}{\partial w_3}r f''(r)\right.\\
&\left.+\frac{\partial \Omega_1}{\partial w_3}[2f^{(3)}(r)+rf^{(4)}(r)]
\right\rbrace,
\end{split}
\end{align}
\begin{align}
\begin{split}
\frac{\partial f''}{\partial w_3}&=\\
&\frac{\Omega_1}{2\left(4 w_1^2\, \Omega_2-\,\Omega_1^2\right)}
\left\lbrace
-2V(r)-2rV'(r)-2\frac{\partial U'(r)}{\partial w_3}
+\frac{\partial \Omega_2}{\partial w_3}[rf'(r)+f(r)]\right.\\
&\left.+\frac{\partial \Omega_1}{\partial w_3}[rf^{(3)}(r)+3f''(r)]
\right\rbrace
+\frac{w_1^2}{4 w_1^2\, \Omega_2-\,\Omega_1^2}
\left\lbrace
-10V''(r)-2rV^{(3)}(r)-2\frac{\partial U^{(3)}(r)}{\partial w_3}\right.\\
&\left.+\frac{\partial \Omega_2}{\partial w_3}[rf^{(3)}(r)+f''(r)]
+\frac{\partial \Omega_1}{\partial w_3}[rf^{(5)}(r)+3f^{(4)}(r)]
\right\rbrace,
\end{split}
\end{align}
\begin{align}
\begin{split}
\frac{\partial f^{(3)}}{\partial w_3}&=\\
&-\frac{\Omega_2}{4 w_1^2\,\Omega_2-\,\Omega_1^2}
\left\lbrace
-2rV(r)-2\frac{\partial U(r)}{\partial w_3}
+\frac{\partial \Omega_2}{\partial w_3}rf(r)
+\frac{\partial \Omega_1}{\partial w_3}[rf''(r)+2f'(r)]
\right\rbrace\\
&-\frac{\Omega_1}{2\left(4 w_1^2\, \Omega_2-\,\Omega_1^2\right)}
\left\lbrace
-8V'(r)-2rV''(r)-2\frac{\partial U''(r)}{\partial w_3}
+\frac{\partial \Omega_2}{\partial w_3}rf''(r)\right.\\
&\left.+\frac{\partial \Omega_1}{\partial w_3}[rf^{(4)}(r)+2f^{(3)}(r)]
\right\rbrace,
\end{split}
\end{align}

\begin{align}
\begin{split}
\frac{\partial f'}{\partial w_1}&=\\
&\frac{\Omega_1}{2\left(4w_1^2\, \Omega_2-\,\Omega_1^2\right)}
\left\lbrace
-2rW(r)-2\frac{\partial U(r)}{\partial w_1}
+\frac{\partial \Omega_2}{\partial w_1}rf(r)
+\frac{\partial \Omega_1}{\partial w_1}[2f'(r)+r f''(r)]\right.\\
&\left.+2w_1[4f^{(3)}(r)+rf^{(4)}(r)]
\right\rbrace \\
&+\frac{w_1^2}{4w_1^2\, \Omega_2-\Omega_1^2}
\left\lbrace
-8W'(r)-2rW''(r)-2\frac{\partial U''(r)}{\partial w_1}
+\frac{\partial \Omega_2}{\partial w_1}r f''(r)\right.\\
&\left.+\frac{\partial \Omega_1}{\partial w_1}[2f^{(3)}(r)+rf^{(4)}(r)]
+2w_1[4f^{(5)}(r)+rf^{(6)}(r)]
\right\rbrace,
\end{split}
\end{align}
\begin{align}
\begin{split}
\frac{\partial f''}{\partial w_1}&=\\
&\frac{\Omega_1}{2\left(4 w_1^2\, \Omega_2-\,\Omega_1^2\right)}
\left\lbrace
-2W(r)-2rW'(r)-2\frac{\partial U'(r)}{\partial w_1}
+\frac{\partial \Omega_2}{\partial w_1}[rf'(r)+f(r)]\right.\\
&\left.+\frac{\partial \Omega_1}{\partial w_1}[rf^{(3)}(r)+3f''(r)]
+2w_1[5f^{(4)}(r)+rf^{(5)}(r)]
\right\rbrace\\
&+\frac{w_1^2}{4 w_1^2\, \Omega_2-\,\Omega_1^2}
\left\lbrace
-10W''(r)-2rW^{(3)}(r)-2\frac{\partial U^{(3)}(r)}{\partial w_1}
+\frac{\partial \Omega_2}{\partial w_1}[rf^{(3)}(r)+f''(r)]\right.\\
&\left.+\frac{\partial \Omega_1}{\partial w_3}[rf^{(5)}(r)+3f^{(4)}(r)]
+2w_1[5f^{(6)}(r)+rf^{(7)}(r)]
\right\rbrace,
\end{split}
\end{align}
\begin{align}
\begin{split}
\frac{\partial f^{(3)}}{\partial w_1}&=\\
&-\frac{\Omega_2}{4 w_1^2\,\Omega_2-\,\Omega_1^2}
\left\lbrace
-2rW(r)-2\frac{\partial U(r)}{\partial w_1}
+\frac{\partial \Omega_2}{\partial w_1}rf(r)
+\frac{\partial \Omega_1}{\partial w_1}[rf''(r)+2f'(r)]\right.\\
&\left.+4w_1[2f^{(3)}(r)+rf^{(4)}(r)]
\right\rbrace\\
&-\frac{\Omega_1}{2\left(4 w_1^2\, \Omega_2-\,\Omega_1^2\right)}
\left\lbrace
-8W'(r)-2rW''(r)-2\frac{\partial U''(r)}{\partial w_1}
+\frac{\partial \Omega_2}{\partial w_1}rf''(r)\right.\\
&\left.+\frac{\partial \Omega_1}{\partial w_1}[rf^{(4)}(r)+2f^{(3)}(r)]
+2w_1[4f^{(5)}(r)+rf^{(6)}(r)]
\right\rbrace.
\end{split}
\end{align}
\end{widetext}

\end{document}